\documentclass[preprint,showpacs,preprintnumbers,amsmath,amssymb,
nofootinbib,eqsecnum,12pt,floatfix]{revtex4}

\usepackage{graphicx}
\usepackage{bm}
\usepackage{eufrak}

\newcommand{\mys}[1]{\section{#1} \setcounter{equation}{0}}

\newlength{\dummysp}
\newcommand{\half}{\frac{1}{2}}

\newcommand{\beq}{\begin{eqnarray}}
\newcommand{\eeq}{\end{eqnarray}}
\newcommand{\nnn}{ \nonumber \\ }
\newcommand{\p}{{\partial}}

\newcommand{\e}{{\epsilon}}
\newcommand{\s}{{\sigma}}
\newcommand{\vev}[1]{{\langle #1 \rangle}}

\newcommand{\ord}[1]{{{\cal O}(#1)}}
\newcommand{\gappeq}{\mathrel{\rlap {\raise.5ex\hbox{$>$}}
{\lower.5ex\hbox{$\sim$}}}}
\newcommand{\lappeq}{\mathrel{\rlap{\raise.5ex\hbox{$<$}}
{\lower.5ex\hbox{$\sim$}}}}
\newcommand{\myref}[1]{(\ref{#1})}

\newcommand{\ben}{\begin{enumerate}}
\newcommand{\een}{\end{enumerate}}

\newcommand{\sqtw}{\sqrt{2}}

\newcommand{\hc}{{\rm h.c.}}

\newcommand{\ddd}{\nnn &&}

\newcommand{\bit}{\begin{itemize}}
\newcommand{\eit}{\end{itemize}}

\newcommand{\susy}{supersymmetry}
\newcommand{\susyc}{supersymmetric}

\newcommand{\Ncal}{{\cal N}}

\newcommand{\Ocal}{{\cal O}}

\newcommand{\phit}{{\tilde \phi}}
\newcommand{\phitb}{{\phit^*}}

\newcommand{\chit}{{\tilde \chi}}

\newcommand{\Ftil}{{\tilde F}}
\newcommand{\Ftilb}{\Ftil^*}
\newcommand{\Fcal}{{\cal F}}

\def\[{\left [}
\def\]{\right ]}
\def\({\left (}
\def\){\right )}

\def\nott#1{\setbox0=\hbox{$#1$}                
   \dimen0=\wd0                                 
   \setbox1=\hbox{/} \dimen1=\wd1               
   \ifdim\dimen0>\dimen1                        
      \rlap{\hbox to \dimen0{\hfil/\hfil}}      
      #1                                        
   \else                                        
      \rlap{\hbox to \dimen1{\hfil$#1$\hfil}}   
      /                                         
   \fi}                                         %

\begin{document}

\author{Chen Chen}
\email{chenc10@rpi.edu}
\author{Eric Dzienkowski}
\email{dziene@rpi.edu}
\author{Joel Giedt}
\email{giedtj@rpi.edu}
\affiliation{Department of Physics, Applied Physics and Astronomy,
Rensselaer Polytechnic Institute, 110 8th Street, Troy NY 12065 USA}

\date{\today}

\begin{abstract}
We numerically evaluate the one-loop counterterms for the four-dimensional
Wess-Zumino model formulated on the lattice using Ginsparg-Wilson
fermions of the overlap (Neuberger) variety, together with an auxiliary
fermion (plus superpartners), such that a lattice version of $U(1)_R$ symmetry is exactly
preserved in the limit of vanishing bare mass.
We confirm previous findings by other authors that at one
loop there is no renormalization of the superpotential in the
lattice theory, but that there is a mismatch in the wavefunction
renormalization of the auxiliary field.
We study the range of the Dirac operator that results when the
auxiliary fermion is integrated out, and 
show that localization does occur, but that it is less pronounced
than the exponential localization of the overlap operator.
We also present preliminary simulation results for this model,
and outline a strategy
for nonperturbative improvement of the lattice supercurrent
through measurements of \susy\ Ward identities.
Related to this, some benchmarks
for our graphics processing unit code are provided.
Our simulation results find a nearly vanishing vacuum
expectation value for the auxiliary field, consistent
with approximate supersymmetry at weak coupling.
\end{abstract}

\title{Lattice Wess-Zumino model with Ginsparg-Wilson fermions: 
One-loop results and GPU benchmarks}



\maketitle

\mys{Introduction}
One would like to have a general method for studying strongly
coupled supersymmetric field theories with lattice techniques.
This is because nonperturbative dynamics play an important
role in the theory of supersymmetry breaking and its transmission
to the visible sector of particle physics.  In this article
we examine one such general method, which involves a fine-tuning
of bare lattice parameters, after having restricted the number
of counterterms using lattice symmetries.  At the same time,
we perform detailed numerical studies of a lattice formulation
that was studied by several groups a few years ago \cite{Fujikawa:2001ka,Fujikawa:2001ns,
Fujikawa:2002ic,Bonini:2004pm,Kikukawa:2004dd,Bonini:2005qx}.  We will
highlight some interesting features of that model and
present some new results regarding locality of the lattice Dirac operator and the
degree of explicit supersymmetry breaking that occurs.

Four-dimensional supersymmetric models on the lattice\footnote{For reviews 
with extensive references see \cite{Giedt:2006pd,Giedt:2007hz,Giedt:2009yd,
Catterall:2009it,Feo:2004kx,Montvay:2001aj}.} generically
require fine-tuning of counterterms.  This is to be contrasted
with lower dimensional theories where lattice symmetries
can eliminate the need for such fine-tuning; see \cite{Catterall:2009it} for further details.
The one known four-dimensional exception is pure $\Ncal=1$ super-Yang-Mills
using Ginparg-Wilson fermions; the domain wall variety has been
the subject of past \cite{Fleming:2000fa} and recent \cite{Giedt:2008cd,Endres:2008tz,Giedt:2008xm,
Endres:2009yp,Giedt:2009yd,Endres:2009pu} simulations.
Clearly we would like to go beyond pure $\Ncal=1$ super-Yang-Mills.
Recently it was proposed \cite{Elliott:2008jp} that an acceptable amount of fine-tuning
could be efficiently performed using a multicanonical Monte Carlo \cite{Berg:1991cf}
simulation together with Ferrenberg-Swendsen reweighting \cite{Falcioni:1982cz,Ferrenberg:1988yz,Ferrenberg:1989ui}
in a large class of theories; see also \cite{Giedt:2009yd}.
In this approach it is necessary to design the
multicanonical reweighting function.  We had suggested beginning at weak
coupling on small lattices, using the one-loop perturbative counterterms
as initial conditions to an iterative search approach.  The
present article reports numerical results for one-loop calculations in lattice perturbation
theory that are designed to locate this starting point.
Ironically we find that for the lattice theory that is
studied here, the one-loop counterterms are entirely wavefunction
renormalization, and that the logarithmically divergent
parts match for the scalar and fermion.  As a result, the
initial condition for the iterative search is equivalent
to starting with the tree-level action, since it is just a
rescaling of the fields.  This lends interest
to our simulation results for the action with no fine-tuning,
which we report here.

The theory that we study is the four-dimensional Wess-Zumino model,
formulated on the lattice with a variant of overlap (Neuberger) fermions \cite{Neuberger:1997fp}.
The goal of the formulation is to impose the Majorana
condition and simultaneously preserve the chiral $U(1)_R$ symmetry \cite{Fujikawa:2001ka,Fujikawa:2001ns,
Fujikawa:2002ic,Bonini:2004pm,Kikukawa:2004dd,Bonini:2005qx}
that is present in the continuum in the massless limit.  As will be
seen, preserving this symmetry greatly limits the number of
counterterms that must be fine-tuned in order to obtain
the supersymmetric continuum limit.  
In addition to overlap fermions, the lattice formulation
has auxiliary fermions (plus superpartner fields) that couple to the overlap fermions
through the Yukawa coupling, as in \cite{Luscher:1998pqa}.  It is possible to integrate
out the auxiliary fermions (and superpartner fields), and when one does this a nonanalytic
Dirac operator results for the surviving fermionic field.
Thus, as has been discussed originally in \cite{Fujikawa:2001ka},
and at greater length in \cite{Fujikawa:2001ns,Kikukawa:2004dd} the action is
singular once auxiliary fields are integrated out.  However,
as we describe below, there is a sensible resolution of
this singularity by taking the theory to ``live'' inside
a finite box, with antiperiodic boundary conditions in the
time direction for the fermions.
The singularity of the Dirac operator that this resolves
is related to nonpropagating modes in the infinite
volume limit; the fact that these are nonpropagating
was shown in \cite{Bonini:2005qx}.  However, singularities
in the Dirac operator raise the spectre of possible nonlocalities
in the continuum limit, as was found in gauge theories with
the SLAC derivative \cite{Karsten:1979wh}.  By analogy to that
study, we have analyzed the continuum limit of the scalar
self-energy and find that it is analytic in $p^2$, so that
there is no sign of nonlocality.  We have also measured the
degree of localization of the Dirac operator following
the approach of \cite{Hernandez:1998et}.  We find that
while there is localization, it is less pronounced than the
exponential localization of the overlap operator.
In the process of discussing our numerical perturbative
results we are able to highlight the divergence structure of this
theory, which turns out to be strictly wave function renormalization
at one-loop.  The wave function renormalization of the
fermion and the physical scalar match at one loop in the continuum
limit of the lattice expressions; but, the auxiliary
scalar has a mismatched wave function counterterm.  These findings appeared
previously in \cite{Fujikawa:2001ka}; thus we confirm those
results.

Having discussed the perturbative results, we then review correlation functions and
renormalization constants that must be measured in
the simulations in order to fine-tune the theory.
These involve the renormalized supercurrent (the current
is renormalized because the symmetry is broken by
the lattice regulator).
We conclude with preliminary simulation results.
In particular, we have developed all of the components
for simulations on graphics processing units.  Benchmarks that
characterize the performance of our code are reported here.  
We measure one broken Ward-Takahashi
identity in a simulation and find it to be very small
at weak coupling.  We argue that this is consistent
with the nonrenormalization of the superpotential at
one loop in lattice perturbation theory.

\section{Definitions}
\subsection{Continuum}
The Euclidean continuum theory has action
\beq
&& S = - \int d^4 x ~ \bigg\{ \half \chi^T C M \chi + \phi^* \Box \phi + F^* F
+ F^* (m^* \phi^* + g^* \phi^{* 2}) + F (m \phi + g \phi^{ 2}) \bigg\},
\ddd M = \nott{\p} + (m + 2 g \phi) P_+ + (m^* + 2 g^* \phi^*) P_- .
\eeq
Our conventions will be ($i=1,2,3$):
\beq
&& \gamma_0 = \begin{pmatrix} 0 & 1 \cr 1 & 0 \end{pmatrix}, \quad
\gamma_i = \begin{pmatrix} 0 & i\s_i \cr -i\s_i & 0 \end{pmatrix}, \quad
\gamma_5 = \begin{pmatrix} -1 & 0 \cr 0 & 1 \end{pmatrix}, \quad
\ddd P_\pm = \half (1 \pm \gamma_5), \quad C = \gamma_0 \gamma_2 .
\eeq
It can be checked that the action is invariant under
the \susy\ transformations
\beq
&&
\delta_\e \phi = \sqtw \e^T C P_+ \chi, \quad \delta_\e \phi^* = \sqtw \e^T C P_- \chi, \quad
\ddd
\delta_\e \chi = - \sqtw P_+ (\nott{\p} \phi + F) \e - \sqtw P_- (\nott{\p} \phi^* + F^*) \e,
\ddd \delta_\e F = \sqtw \e^T C \nott{\p} P_+ \chi, \quad 
\delta_\e F^* = \sqtw \e^T C \nott{\p} P_- \chi
\eeq
Note also that in momentum space
\beq
- \phi^* \Box \phi \to |\phi(p)|^2 p^2
\eeq
so that for a well defined partition function we must take
the negative sign in the exponent:
\beq
Z = \int [d\phi ~ d\phi^* ~ dF ~ dF^* ~ d\chi] ~~ e^{-S}
\eeq
On the other hand, integration over the auxiliary
field $F$ is not well-defined (as is usual in Euclidean
formulations of \susyc\ theories).  In particular,
we have after integrating out the fermions and completing the square,
\beq
Z &=& \int [d \phi ~ d \phi^*] e^{-S(\phi)} {\rm Pf}CM(\phi) \int [d F ~ d F^*] \exp \int d^4 x ~ |F^* + m \phi + g \phi^2|^2,
\nnn S(\phi) &=& \int d^4x ~ \( - \phi^* \Box \phi + |m \phi + g \phi^2|^2 \)
\label{tfi}
\eeq
where ${\rm Pf}CM$ is the Pfaffian of the matrix $CM$.
Under shift of integration variables
\beq
F \to F' = F + m^* \phi^* + g^* \phi^{*2}, \quad
F^* \to F'^* = F^* + m \phi + g \phi^2
\eeq
$Z$ does not converge, due to the auxiliary field factor $\int [d F' ~ d F'^*] \exp \int d^4 x ~ |F'|^2$.
This fact was noted, for example, in \cite{Fujikawa:2001ka} where they simply divide
by an infinite constant---the partition function of the $g=0$ theory---to
cancel the infinity.  In a lattice simulation we do not
have this luxury, and we must decide on the proper way
to deal with the integration over $F,F^*$.

Of course in the Minkowski space formulation, due to the
presence of an additional factor $-i$, the integral over
the auxiliary fields $F',F'^*$ can be computed by
analytic continuation.  One discards the overall
constant into the normalization of $Z$, and the
net result is that one imposes the equations of motion:
\beq
\frac{\delta S}{\delta F(x)} = \frac{\delta S}{\delta F^*(x)} = 0.
\label{feom}
\eeq
From this we conclude that
in Euclidean space the integral \myref{tfi} should be understood
in a formal sense; it is an instruction to impose \myref{feom}.
This is equivalent to completing the square, as in \myref{tfi},
shifting the integration variable and then ignoring the (infinite) constant
factor that is generated.  Thus all simulations are performed with
the action in the form with the auxiliary fields removed.
In this form the \susy\ transformations
are nonlinear and the \susy\ algebra only closes on-shell.

\subsection{Lattice}
We next discuss the lattice action, which is a special case of 
the formulations of \cite{Fujikawa:2001ka,Fujikawa:2001ns}; we
write the lattice action in forms given in
\cite{Bonini:2004pm,Kikukawa:2004dd,Bonini:2005qx}.
For this, a few lattice derivative operators must be introduced.
\beq
A &=& 1 - a D_W, \quad D_W = \half \gamma_\mu ( \p_\mu^* + \p_\mu) + \half a \p_\mu^* \p_\mu
\nnn
D_1 &=& \half \gamma_\mu ( \p_\mu^* + \p_\mu) (A^\dagger A)^{-1/2}
\nnn
D_2 &=& \frac{1}{a} \[ 1 - \( 1 + \half a^2 \p_\mu^* \p_\mu \) (A^\dagger A)^{-1/2} \]
\nnn
D &=& D_1 + D_2 = \frac{1}{a} \( 1 - A (A^\dagger A)^{-1/2} \)
\eeq
where $\p_\mu$ and $\p_\mu^*$ are the forward and backward difference
operators respectively.  Then the lattice action is \cite{Kikukawa:2004dd}:
\beq
S & = & -a^4 \sum_x \bigg\{ \half \chi^T C D \chi
+ \phi^* D_1^2 \phi  + F^* F + F D_2 \phi + F^* D_2 \phi^*
\ddd - \frac{1}{a} X^T C X - \frac{2}{a} \( \Fcal \Phi + \Fcal^* \Phi^* \)
\ddd + \half \chit^T C \( mP_+ + m^* P_- + 2 g \phit P_+ + 2 g^* \phitb P_- \) \chit
\ddd + \Ftilb (m^* \phitb + g^* \phit^{*2}) + \Ftil (m \phit + g \phit^{ 2}) \bigg\}
\label{kisu}
\eeq
Here, the tilded fields are the linear combinations
\beq
\phit = \phi + \Phi, \quad \chit = \chi + X, \quad \Ftil = F + \Fcal
\eeq
The fields $\Phi, X, \Fcal$ and their conjugates are auxiliary
fields introduced to allow for a lattice realization of the chiral
$U(1)_R$ symmetry in the $m \to 0$ limit:
\beq
&& \delta \chi = i \alpha \gamma_5 \( 1 - \frac{a}{2} D \) \chi + i \alpha \gamma_5 X,
\quad \delta X = i \alpha \gamma_5 \frac{a}{2} D \chi,
\ddd \delta \phi = -3i \alpha \phi + i \alpha \[  \( 1-\frac{a}{2} D_2 \) \phi
- \frac{a}{2} F^* \] + i \alpha \Phi,
\ddd
\delta \Phi = -3i\alpha \Phi + i \frac{a}{2} \alpha \( D_2 \phi + F^* \)
\ddd
\delta F = 3i\alpha F + i \alpha \[ \( 1 - \frac{a}{2} D_2 \) F - \frac{a}{2} D_1^2 \phi^* \]
+ i \alpha \Fcal
\ddd
\delta \Fcal = 3i\alpha \Fcal + i \frac{a}{2} \alpha \( D_2 F + D_1^2 \phi^* \)
\eeq
which takes a particularly simple form on the tilded variables:
\beq
\delta \chit = i \alpha \gamma_5 \chit, \quad \delta \phit = - 2 i \alpha \phit, 
\quad \delta \Ftil = 4 i \alpha \Ftil
\label{tiltra}
\eeq

The \susy\ transformations of the untilded fields will be taken as
\beq
&&
\delta_\e \phi = \sqtw \e^T C P_+ \chi, \quad \delta_\e \phi^* = \sqtw \e^T C P_- \chi,
\ddd
\delta_\e \chi_\beta = - \sqtw (P_+ (D_1 \phi + F) \e)_\beta - \sqtw (P_- (D_1 \phi^* + F^*) \e)_\beta,
\ddd \delta_\e F = \sqtw \e^T C D_1 P_+ \chi, \quad 
\delta_\e F^* = \sqtw \e^T C D_1 P_- \chi
\label{stuntil}
\eeq
and for the auxiliary multiplet
\beq
&&
\delta_\e \Phi = \sqtw \e^T C P_+ X, \quad \delta_\e \Phi^* = \sqtw \e^T C P_- X,
\ddd
\delta_\e X_\beta = - \sqtw (P_+ (D_1 \Phi + \Fcal) \e)_\beta - \sqtw (P_- (D_1 \Phi^* + \Fcal^*) \e)_\beta,
\ddd \delta_\e \Fcal = \sqtw \e^T C D_1 P_+ X, \quad 
\delta_\e \Fcal^* = \sqtw \e^T C D_1 P_- X
\label{staux}
\eeq
Of course this is not a symmetry of the lattice action for $g \not= 0$.
Here we are just choosing the form of the tranformation that will
be used to generate broken Ward-Takahashi identities on the lattice.
The \susy\ transformations of the tilded fields are
\beq
&&
\delta_\e \phit = \sqtw \e^T C P_+ \chit, \quad \delta_\e \phit^* = \sqtw \e^T C P_- \chit,
\ddd
\delta_\e \chit_\beta = - \sqtw (P_+ (D_1 \phit + \Ftil) \e)_\beta - \sqtw (P_- (D_1 \phit^* + \Ftil^*) \e)_\beta,
\ddd \delta_\e \Ftil = \sqtw \e^T C D_1 P_+ \chit, \quad 
\delta_\e \Ftil^* = \sqtw \e^T C D_1 P_- \chit
\label{sttil}
\eeq

As noted in \cite{Kikukawa:2004dd}, we can integrate out the auxiliary
fields $X,\Phi,\Fcal$, treating the tilded fields as
constant, to obtain the lattice action:\footnote{Integrating out
an auxiliary fermion to obtain the fermionic part of this action
was previously noted in \cite{Fujikawa:2001ns}.  There it was noted that this relates
the fermionic action to the one of \cite{Luscher:1998pqa} by
a singular field transformation.  This singularity will be discussed more
in Section \ref{moan} below.}
\beq
S& =& -a^4 \sum_x \bigg\{ \half \chit^T C M \chit
- \frac{2}{a} \phitb D_2 \phit  + \Ftilb (1 - \frac{a}{2} D_2)^{-1} \Ftil \ddd
+ \Ftilb (m^* \phitb + g^* \phit^{*2}) + \Ftil (m \phit + g \phit^{ 2}) \bigg\} .
\label{lata}
\eeq
This is the lattice action Eq.~(2.14) of \cite{Bonini:2004pm} with a notation that
interchanges $D_1 \leftrightarrow D_2$,
which is equivalent to Eq.~(2.22) of \cite{Fujikawa:2001ns} for the $k=0$ case,
using the identities\footnote{We thank A.~Feo for explaining this point 
and providing us with a derivation
of these relations.}
\beq
\Gamma_5 = \gamma_5 (1 - \frac{a}{2} D), \quad \Gamma_5^2 = 1 - \frac{a}{2} D_2,
\quad D^\dagger D = \frac{2}{a} D_2 .
\eeq
The fermion matrix is:
\beq
M = \nott{D} + mP_+ + m^* P_- + 2 g \phit P_+ + 2 g^* \phitb P_- ,
\quad \nott{D} = (1 - \frac{a}{2} D_2)^{-1} D_1
\label{fmat}
\eeq
This way of writing the Dirac operator can be related to the
one that appears in \cite{Fujikawa:2001ns} by the identity:
\beq
(1 - \frac{a}{2} D_2)^{-1} D_1 = (1 - \frac{a}{2} D)^{-1} D
\eeq
Furthermore we can integrate out the auxiliary
fields $\Ftil, \Ftilb$ to obtain the action
\beq
S& =& a^4 \sum_x \bigg\{ -\half \chit^T C M \chit
+ \frac{2}{a} \phitb D_2 \phit  
+ (m^* \phitb + g^* \phit^{*2}) (1 - \frac{a}{2} D_2) (m \phit + g \phit^{ 2}) \bigg\}
\label{acws}
\eeq
This is the action that is used in our numerical simulations.

When fine-tuning of the lattice action is performed,
we must invoke the most general lattice action consistent
with symmetries.  Since we perform our simulations at $m \not= 0$,
this is just the action with all dimension $\leq 4$ operators
built out of the physical fields, $\phit$ and $\chit$.
We write it here for reference:
\beq
S &=&  a^4 \sum_x \bigg\{ -\half \chit^T C ( \nott{D} + m_1 P_+ + m_1^* P_- ) \chit
+ \frac{2}{a} \phitb D_2 \phit  
\ddd
+ m_2^2 |\phit|^2 + \lambda_{1} |\phit|^4
+ \big( m_3^2 \phit^2 + g_1 \phit^3 
+ g_2 \phit \phit^{*2} + \lambda_2 \phit^4 + \lambda_{3} \phit \phit^{*3} + \hc \big)
\ddd
- \chit^T C ( y_1 \phit P_+ + y_1^* \phit^* P_- ) \chit
- \chit^T C ( y_2 \phit P_- + y_2^* \phit^* P_+ ) \chit \bigg\}
\label{mogeac}
\eeq
A term linear in $\phit$ has been eliminated
by the redefinition $\phit \to \phit + c$ with $c$ a constant.  The parameters
$m_2^2$ and $\lambda_1$ are real and all other parameters
are complex.  Whereas in
the supersymmetric theory there are four real parameters,
in the most general theory we have eighteen real parameters to
adjust.  Holding $m_1$ and $y_1$ fixed, we have fourteen 
real parameters that must be adjusted to obtain the supersymmetric
limit.  The counting can be alleviated somewhat by imposing
CP invariance, so that all parameters can be assumed real.
Then we have a total of ten parameters.  Holding two fixed,
we must tune the other eight to achieve the supersymmetric limit.
Conducting a fine-tuning in an eight-dimensional parameter
space is a daunting task.

On the other hand in the limit $m_1 \to 0$ we can impose the
$U(1)_R$ symmetry \myref{tiltra}.  This restricts the action
to
\beq
S &=&  a^4 \sum_x \bigg\{ -\half \chit^T C \nott{D} \chit
+ \frac{2}{a} \phitb D_2 \phit  
+ m_2^2 |\phit|^2 + \lambda_{1} |\phit|^4
\ddd
- \chit^T C ( y_1 \phit P_+ + y_1^* \phit^* P_- ) \chit
\bigg\}
\label{symgeac}
\eeq
If we hold $y_1$ fixed, then only $m_2^2$ and $\lambda_1$ must be
fine-tuned.  Conducting a search in a two-dimensional parameter
space, with both coming from bosonic terms, is manageable.
The difficult part is that we must extrapolate to the massless
fermion limit.  Another potential problem is that we impose
antiperiodic boundary conditions for the fermion in the time
direction, but must impose periodic boundary conditions for
the scalar in order for the action to be single-valued 
on the circle in the time direction.  This breaks supersymmetry explicitly by boundary
conditions.  At finite mass this should be an effect that
can be made arbitrarily small by taking the large volume limit.
However at vanishing mass, there will be long distance modes
that will ``feel'' the breaking due to boundary conditions.\footnote{We
thank G.~Bergner for raising this point.}  Thus it is important
that we take $T \gg 1/ma$ as $m$ is sent to zero, where $T$ is
the number of sites in the time direction.

As we will see, the one-loop behavior of
the theory \myref{lata} closely follows that of the continuum, so that no new
operators are generated at this order.  Thus at this level
of approximation, a fine-tuning of the general lattice action \myref{mogeac} is not needed.
For reasons that will be clearer once we have presented our
perturbative results, it is of interest to study the original lattice
action \myref{acws} in our simulations, without any fine-tuning.  By
measuring the degree of \susy\ breaking through \susy\ Ward
identities that are conserved in the continuum, we gain
information about the higher orders and nonperturbative
aspects of the lattice theory.

\section{Modes, Feynman rules and tilde/untilde equivalence}
\label{moan}
\subsection{Mode analysis}
As to the the nonlocal operator $(1 - \half a D_2)^{-1}$
that appears in the action \myref{lata}, note that for smooth
field configurations $a D_2 \sim a^2 \Box$.
Nevertheless, one may rightly worry about the effect of modes for which $D_2 \sim 1/a$.
The spectrum of $1 - \half a D_2$ can easily be calculated in momentum
space:
\beq
(1 - \half a D_2)(p) = \half + \half \frac{ 1 - 2 \sum_\mu \sin^2(p_\mu a/2 )}{
\sqrt{ [ 1 - 2 \sum_\mu \sin^2(p_\mu a/2 )]^2  + \sum_\mu \sin^2(p_\mu a)}}
\eeq
One sees that there are zeros at $p_\mu a$ equal to all of
the would-be doublers:
\beq
(\underline{\pi, 0, 0, 0}), \quad (\underline{\pi, \pi, 0, 0 } ), \quad
(\underline{\pi, \pi, \pi, 0}), \quad (\pi,\pi,\pi,\pi),
\label{woube}
\eeq
where the underline indicates all possible permutations.  Thus 
the auxiliary field kinetic term appearing in \myref{lata} is singular, as was
discussed in \cite{Fujikawa:2001ka}, and more
at length in \cite{Fujikawa:2001ns,Kikukawa:2004dd,Bonini:2005qx}.  For the fermions
we must also consider
\beq
D_1(p) = \frac{-i a^{-1} \sum_\mu \gamma_\mu \sin(p_\mu a)}{
\sqrt{ [ 1 - 2 \sum_\mu \sin^2(p_\mu a/2 )]^2  + \sum_\mu \sin^2(p_\mu a)}}
\label{d1mom}
\eeq
Thus $D_1(p)=0$ for the would-be doublers \myref{woube} and
the fermion operator $(1 - \half a D_2)^{-1} D_1$ is indeterminate.
In \cite{Bonini:2005qx} it was shown that in a limiting
process of approaching these points in momentum space, $(1 - \half a D_2)^{-1} D_1$
is divergent.  Thus the would-be doublers are actually
nonpropagating.  Note that
\beq
(1 - \half a D_2)^{-1} D_1 \gamma_5 + \gamma_5 (1 - \half a D_2)^{-1} D_1  = 0
\eeq
which is how the lattice formulation manages to preserve
$U(1)_R$ symmetry.  This is not in conflict with the Nielsen-Ninomiya
theorem \cite{Nielsen:1981hk,Nielsen:1980rz} precisely because the operator $(1 - \half a D_2)^{-1} D_1$
is unbounded and therefore nonanalytic.  One should then worry about locality since
a nonanalytic Dirac operator might violate this property,
as was true of the SLAC Dirac operator \cite{Karsten:1979wh}.
We will investigate this below.  We will find numerically
that it does show localization, though not as pronounced
as the exponential localization of the overlap operator.
In fact it has a long tail that is a cause of some concern.
However our perturbative results in a later section
show no sign of nonanalytic sickness in the scalar
self-energy as a function of $p^2$, which would be the
analogue of the results of \cite{Karsten:1979wh}.

For the fermion, in our simulations, we address the difficulty of the indeterminate
operator by taking the fermion field to be antiperiodic in the time
direction and restrict considerations to finite lattices $L^3 \times T$
with $T$ even.  Taking $T$ even is necessary in order
to avoid other zeros of $(1 - \half a D_2)$.  This resolves the
indeterminacies for all finite $L,T$ with $T$ even.  One then defines the infinite
spacetime volume theory rigorously as the limiting value $L,T \to \infty$.
Unfortunately, the maximum eigenvalue
of the fermion operator $(1 - \half a D_2)^{-1} D_1$ diverges as $T \to \infty$,
so that the fermion matrix becomes poorly conditioned at very large system
size.  For the $T=32$ and $T=64$ lattices that we study this does not
prove to be a serious problem.

For the auxiliary field $\Ftil$, in our simulations, we formally integrate it out,
leading to the action \myref{acws}.  For the modes with $(1 - \frac{a}{2} D_2)(p)=0$
the scalar potential vanishes.  However, the kinetic term $\frac{2}{a} \phitb D_2 \phit$
is nonvanishing on this modes, with $\frac{2}{a} D_2 (p) = 4 / a^2$.  These
modes are therefore suppressed in the path integral and are part of
the cutoff theory.

In perturbation theory, one can arrange for the modes at $p_\mu a$ of
the would-be doublers \myref{woube} to be nonpropagating, as
ought to be based on the results of \cite{Bonini:2005qx}. 
This nonpropagating feature is
incorporated into the Feynman diagram rules of \cite{Kikukawa:2004dd},
as we now explain.  First we note that
\beq
&& \[ - D_1 + \( 1 - \frac{a}{2} D_2 \) (m^* P_+ + m P_-) \]
\[ \( 1 - \frac{a}{2} D_2 \)^{-1} D_1 + (m P_+ + m^* P_-) \]
\ddd = \frac{2}{a} D_2 + |m|^2 \( 1 - \frac{a}{2} D_2 \).
\eeq
using the Ginsparg-Wilson relation, which in terms of $D_1$ and $D_2$
takes the form \cite{Bonini:2004pm}
\beq
D_1^2 - D_2^2 = - \frac{2}{a} D_2
\eeq
Then we find that the fermion propagator can be written
\beq
\vev{\chit(x) \chit^T(y)} C = a^{-4}
\( \frac{ - D_1 + \( 1 - \frac{a}{2} D_2 \) (m^* P_+ + m P_-)}
{\frac{2}{a} D_2 + |m|^2 \( 1 - \frac{a}{2} D_2 \)} \)(x,y)
\label{ksprop}
\eeq
This propagator vanishes for the would-be doublers \myref{woube}.
It avoids the alternative form that comes from a straightforward
inversion of the free Dirac operator $M_0 = \( 1 - \frac{a}{2} D_2 \)^{-1} D_1 + m P_+ + m^* P_-$:
\beq
\vev{\chit(x) \chit^T(y)} C = -a^{-4}
\( \frac{\( 1 - \frac{a}{2} D_2 \)^{-1} D_1 - (m^* P_+ + m P_-)}
{ \frac{2}{a} D_2 \( 1 - \frac{a}{2} D_2 \)^{-1} + |m|^2 } \) (x,y)
\eeq
which is indeterminate for the would-be doublers.
In our perturbative analysis of Section \ref{s:onel} below,
we use the propagator \myref{ksprop}.  This allows us
to use periodic boundary conditions for the one-loop calculations.

\subsection{Locality}
\begin{figure}
\begin{center}
\includegraphics[width=3in,height=5in,angle=90]{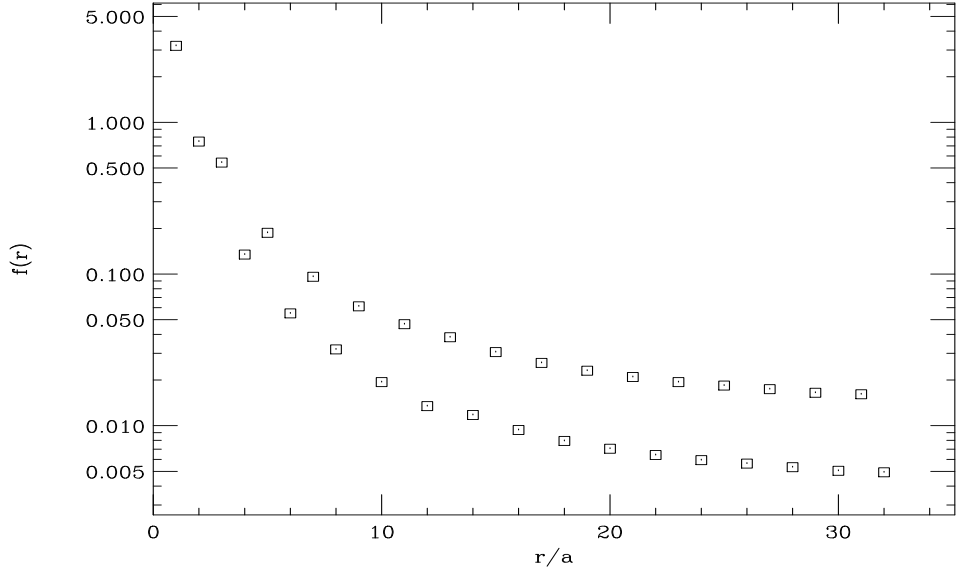}
\caption{Probe of the range of the $\nott{D}$ operator, Eq.~\myref{uir}.  The value of $f$ at
$r=0$ is zero and is not shown in this plot.  Even $r$ are consistently
lower than odd $r$.  \label{rfig}}
\end{center}
\end{figure}

Following the numerical approach of \cite{Hernandez:1998et} we
can investigate the locality of the operator $\nott{D} = \( 1 - \frac{a}{2} D_2 \)^{-1} D_1$.
In what follows, we impose the antiperiodic in time boundary
conditions on the fermions.
We introduce a unit point source $\eta$ at the origin $x=0$
and then compute
\beq
\psi(x) =  \[ \( 1 - \frac{a}{2} D_2 \)^{-1} D_1 \eta \] (x)
\eeq
The ``taxi-driver distance''
\beq
r = ||x||_1 = \sum_\mu ( |x_\mu| \; \text{or} \; |L_\mu - x_\mu| )
\eeq
to the origin is defined; the shortest length is selected in each ``or''
statement, where $L_\mu/a$ is the number of lattice sites
in the $\mu$ direction.  One then computes the norm $||\psi(x)||$ in spinor index space
at site $x$ and obtains the function
\beq
f(r) = \max \{ ||\psi(x)|| ~~ | ~~ ||x||_1 = r \}
\label{uir}
\eeq
We show this function for a $16^4$ lattice in Fig.~\ref{rfig}.
What one sees is that there is a long tail on the operator.
In contrast to the overlap operator, we do not find exponential localization;
the localization that does occur is less pronounced and
we view it as an open question whether or not this is harmful.

\subsection{Tilde/untilde equivalence}
Suppose we wish to compute correlation
functions of the untilded fields.  For this purpose we introduce
sources that couple only to them:
\beq
S_{src} = -a^4 \sum_x \bigg\{ \eta^T C P_+ \chi + \tau^T C P_- \chi
+ J^* \phi + J \phi^* + k^* F + k F^* \bigg\}
\label{Ssrc}
\eeq
Next we integrate out the auxiliary fields $X, \Phi, \Fcal$.
The corresponding equations of motion, determined by variation of
$X, \Phi, \Fcal$ holding $\tilde\chi, \tilde\phi, \tilde F$ constant, involve the sources.
These can then be solved for the auxiliary fields $X, \Phi, \Fcal$:
\beq
X &=& -\frac{a}{2} \( 1 - \frac{a}{2} D \)^{-1} \[ D \chit + P_+ \eta + P_- \tau \] \nnn
\Phi &=& -\frac{a}{2} \( 1 - \frac{a}{2} D_2 \)^{-1} {\tilde F}^* - \frac{a}{2} k^*
- \frac{a^2}{4} \( 1 - \frac{a}{2} D_2 \)^{-1} J \nnn
\Fcal &=& D_2 \phit^* + \frac{a}{2} D_2 k - \frac{a}{2} J
\eeq
Thus when these are used to eliminate $X, \Phi, \Fcal$,
quadratic terms in the sources are generated.  We obtain
\beq
S + S_{src} &=& -a^4 \sum_x \bigg\{ \half \chit^T C \( 1 - \frac{a}{2} D \)^{-1} D \chit
- \frac{2}{a} \phit^* D_2 \phit + \Ftil^* \( 1 - \frac{a}{2} D_2 \)^{-1} \Ftil
\ddd + \half \chit^T \( mP_+ + m^* P_- + 2 g \phit P_+ + 2 g^* \phitb P_- \) \chit
\ddd + \Ftilb (m^* \phitb + g^* \phit^{*2}) + \Ftil (m \phit + g \phit^{ 2})
\ddd + \frac{a}{4} (\eta^T P_+ + \tau^T P_- ) C \( 1 - \frac{a}{2} D \)^{-1} (P_+ \eta + P_- \tau)
\ddd + (\eta^T P_+ + \tau^T P_- ) C \( 1 - \frac{a}{2} D \)^{-1} \chit
+ k (\Ftil^* - D_2 \phit + \frac{a}{2} J) 
\ddd + k^* (\Ftil - D_2 \phit^* + \frac{a}{2} J^*)  
+ J \phit^* + J^* \phit - \frac{a}{2} k^* D_2 k 
\ddd + \frac{a}{2} \Ftil \( 1 - \frac{a}{2} D_2 \)^{-1} J
+ \frac{a}{2} \Ftil^* \( 1 - \frac{a}{2} D_2 \)^{-1} J^*
- \frac{a^2}{4} ( JJ + J^* J^*)
\ddd + \frac{a^2}{4} J^* \( 1 - 2 a D_2 + \frac{a^2}{4} ( D_1^2 + D_2^2 ) \) 
\( 1 - \frac{a}{2} D_2 \)^{-2} J
\bigg\}
\eeq
As far as the source terms for the elementary fields $\chi$ and $\phi$ are concerned,
working with the tilded fields is the same as working with the untilded
fields up to $\ord{a}$ corrections.  Note however that
the source terms involving $k,k^*$ have unusual
terms that do not vanish in the continuum limit.
The implication is that correlation functions for $F,F^*$ differ
from those of $\Ftil,\Ftil^*$ in a way that is guaranteed not
to vanish in the continuum limit.

To see what are the implications for the physical fields,
we set $k=k^*=0$ and integrate out the auxiliary fields $\Ftil,\Ftil^*$:
\beq
S + S_{src} &=& -a^4 \sum_x \bigg\{ \half \chit^T C \( 1 - \frac{a}{2} D \)^{-1} D \chit
- \frac{2}{a} \phit^* D_2 \phit
\ddd + \half \chit^T \( mP_+ + m^* P_- + 2 g \phit P_+ + 2 g^* \phitb P_- \) \chit
\ddd + (m^* \phitb + g^* \phit^{*2}) \( 1 - \frac{a}{2} D_2 \) (m \phit + g \phit^{ 2})
\ddd + \frac{a}{4} (\eta^T P_+ + \tau^T P_- ) C \( 1 - \frac{a}{2} D \)^{-1} (P_+ \eta + P_- \tau)
\ddd + (\eta^T P_+ + \tau^T P_- ) C \( 1 - \frac{a}{2} D \)^{-1} \chit
\ddd + \[ J^* \( \phit - \frac{a}{2} (m \phit + g \phit^{ 2}) \) + \hc \]
- \frac{a^2}{4} ( J^2 + J^{*2})
\ddd - \frac{a^3}{2} J^* \(D_2 - \frac{a^2}{8} ( D_1^2 + D_2^2 ) \) 
\( 1 - \frac{a}{2} D_2 \)^{-2} J
\bigg\}
\label{snoa}
\eeq
To interpret this result,
suppose instead we had introduced sources $\tilde J, {\tilde J}^*, \tilde \eta, \tilde \tau$
for the tilded fields $\phit^*, \phit, P_+ \chit, P_- \chit$: 
\beq
S_{src} = -a^4 \sum_x \bigg\{ {\tilde\eta}^T C P_+ \chit + {\tilde\tau}^T C P_- \chit
+ {\tilde J}^* \phit + {\tilde J} \phit^*  \bigg\}
\eeq
From \myref{snoa} what we see is that
\beq
\frac{\delta}{\delta J(x)} = \frac{\delta}{\delta \tilde J(x)} + \ord{a}
\eeq
and likewise for the other sources ${\tilde J}^*, \tilde \eta, \tilde \tau$.  This is another way of saying
that the correlation functions for the tilded fields are equal to
the correlation functions of the untilded fields, up to $\ord{a}$ corrections.
Of course if there were $\ord{1/a}$ or worse divergences in the
correlation functions, then the difference between the two
sets of correlation functions does not vanish in the continuum limit.
However, what we will find below is that at one loop the
divergences are only $\ln a$, hence the tilded and untilded
correlation functions become equal in the continuum limit.
If $\ord{1/a}$ or worse divergences appear at higher
orders, then the two formulations are not equivalent in
the continuum limit.  In that case one is free to choose
one or the other as the subject for fine-tuning in order
to approach the continuum theory.  We will choose the tilded
formulation in our work.

\subsection{Naive continuum limit}
For reference, we state the basic properties of the
lattice theory the guarantee that the correct continuum
limit is achieved classically.
Two useful identities involving the lattice derivative
operators are
\beq
\lim_{a \to 0} D_1 = \nott{\p}, \quad
\lim_{a \to 0} \frac{2}{a} D_2 = -\Box
\eeq
Then in the naive continuum limit
\beq
(1 - \frac{a}{2} D_2)^{-1} \to (1 + \frac{a^2}{4} \Box)^{-1} \to 1
\eeq
Thus we see that for the kinetic terms
\beq
&&a^4 \sum_x \bigg\{ \half \chit^T C (1 - \frac{a}{2} D_2)^{-1} D_1 \chit
- \frac{2}{a} \phitb D_2 \phit  + \Ftilb (1 - \half a D_2)^{-1} \Ftil \bigg\}
\ddd \to \int d^4x ~ \bigg\{ \half \chit^T C \nott{\p} \chit
+ \phitb \Box \phit  + \Ftilb \Ftil \bigg\}
\eeq
It is interesting that the continuum limit of the action
written in terms of the tilded fields is correct.  This
is a further indication that we can just as well treat them as the ``physical''
variables and work entirely in terms of the action \myref{lata}.

\section{One-loop calculation}
\label{s:onel}
In \cite{Fujikawa:2001ka} an identical calculation is performed,
in that they also compute one-loop corrections to the propagators
and proper vertices.  The action that they use, described by their Eq.~(2.17),
is different from the one employed here, but is related
by a singular field transformation \cite{Fujikawa:2001ns}.  We have
confirmed the results of \cite{Fujikawa:2001ka}, working
in this alternative but perturbatively equivalent formulation.  Our results also
lead to conclusions identical to those of \cite{Kikukawa:2004dd},
who take a different approach to studying the same action
as the one we use, our Eq.~\myref{lata}; they study
the counterterms that must be added to the action
in order for the restoration of the \susy\ Ward-Takahashi identity
to occur.

\subsection{Definitions}
Throughout our presentation, we make use of
\beq
P_\mu(k) \equiv a^{-1} \sin(k_\mu a), \quad
Q_\mu(k) \equiv 2 a^{-1} \sin( k_\mu a/2 ),
\eeq
periodic functions that reduce to momentum in the naive continuum limit.
It is also useful to define
\beq
M(k) = 1 - 2 \sum_\mu \sin^2(k_\mu a/2 ) = 1 - \frac{a^2}{2} Q^2(k),
\quad
s(k) = \sqrt{P^2(k)a^2+M^2(k)}
\eeq
which has the property $\lim_{a \to 0} s(k) = 1$.

\subsection{Analytic results}
\begin{figure}
\begin{center}
\includegraphics[width=5in,height=3in,clip,bb=100 500 500 700]{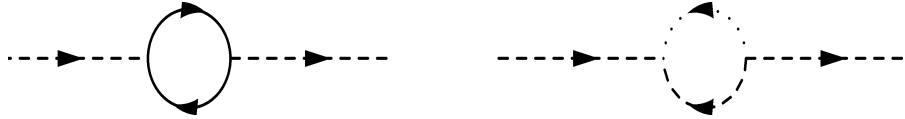}
\caption{Scalar 2-point function that contributes to self-energy. \label{ccd1}}
\end{center}
\end{figure}

The only nonvanishing one-loop scalar two-point function is
the one that gives a one-loop correction to the
nonholomorphic term in the effective action,
\beq
\int d^4x ~ d^4y ~ \phit^*(x) G_{\phit^* \phit}^{-1}(x,y) \phit(y)
\eeq
where $G(x,y)$ is the scalar propagator.  It is obtained from the
sum of the two diagrams shown in Fig.~\ref{ccd1}, yielding:
\beq
a^{2} |g|^2 \int_{-\pi/a}^{\pi/a} \frac{d^4 k}{(2\pi)^4} \frac{N(k,p)}{D(k,p)}, \quad
N(k,p)=\frac{n(k,p)}{d(k,p)} - t(k)
\eeq
\beq
n(k,p) = a^2 P(p+k) \cdot P(k) , \quad d(k,p) = s(k)s(k+p),
\quad t(k) = \frac{ P^2(k)a^2 }{s^2(k)}
\eeq
\beq
D(k,p)= r(k)r(k+p), \quad
r(k) = 2\( 1 - \frac{M(k)}{s(k)} \)
+ \frac{|m|^2 a^2}{2} \( 1 + \frac{M(k)}{s(k)} \)
\eeq
The result vanishes at $p=0$, because $N(k,0)=0$, so no mass counterterm
arises from this diagram.

It is interesting that a term in the effective
action of the form
\beq
\int d^4 x ~ d^4 y ~ \phit(x) G_{\phit \phit}^{-1}(x,y) \phit(y)
\eeq
is not generated at one-loop, since it is allowed by
the symmetries of the lattice action
if $m \not= 0$ and $g \not= 0$.  At one-loop there is an exact
cancellation between the scalar and fermion loops
for all external momentum $p$.

\begin{figure}
\begin{center}
\includegraphics[width=3in,height=1.5in,clip,bb=100 600 300 700]{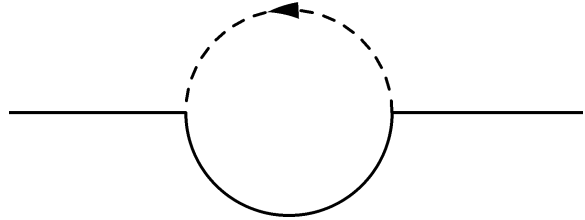}
\caption{Fermion 2-point function that contributes to self-energy. \label{ccd2}}
\end{center}
\end{figure}

\begin{figure}
\begin{center}
\includegraphics[width=3in,height=1.5in,clip,bb=100 600 300 700]{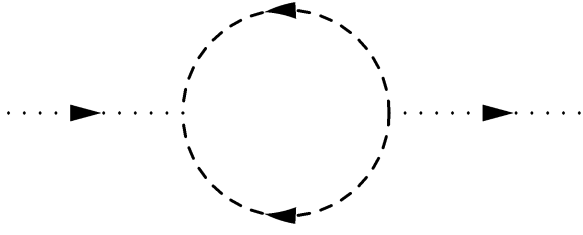}
\caption{Auxiliary 2-point function that contributes to self-energy. \label{ccd3}}
\end{center}
\end{figure}

The diagram that corrects  the fermion propagator, Fig.~\ref{ccd2}, is given by:
\beq
(-i) |g|^2 \int\frac{a^4 d^4k}{(2\pi)^4}P_-
\frac{\nott{P}(p+k)}{r(k)r(p+k)s(p+k)}
\eeq
The diagram Fig.~\ref{ccd3} gives the correction to the $\Ftilb \Ftil$
term and is given by:
\beq
(-1)|g|^2\int\frac{a^4 d^4k}{(2\pi)^4}\frac{1}{r(k)r(p+k)}
\eeq

\begin{figure}
\begin{center}
\includegraphics[width=3in,height=2in,clip,bb=100 550 300 700]{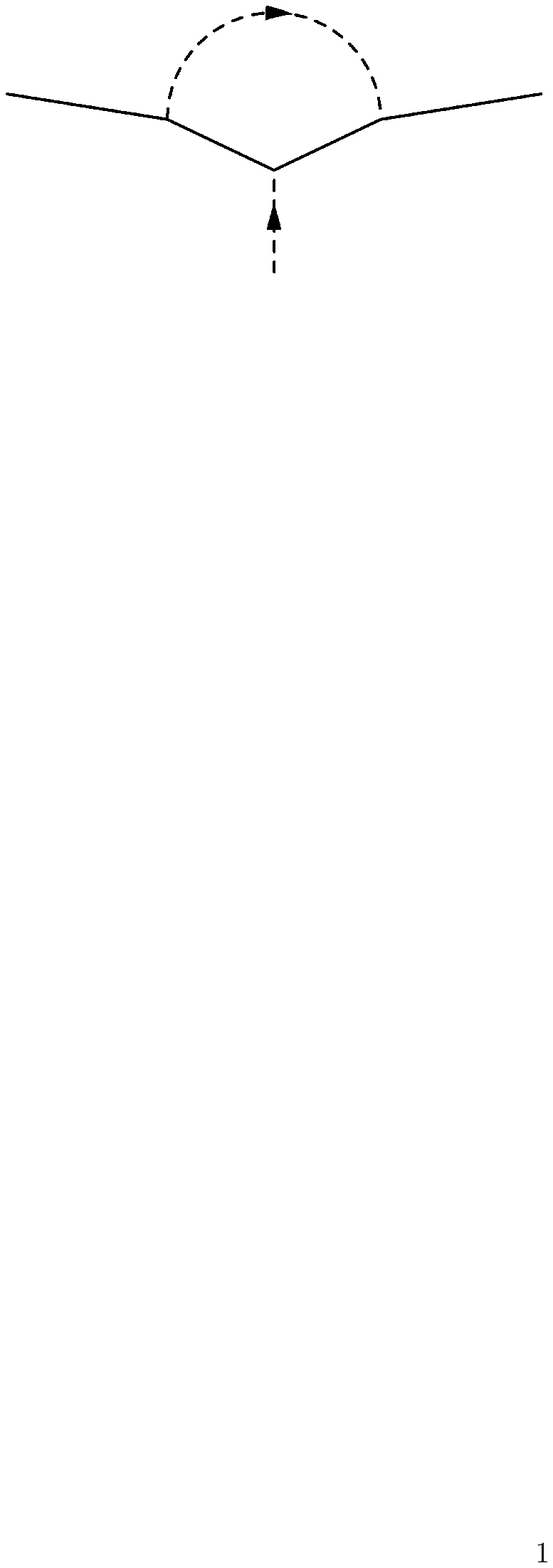}
\caption{3-point function. \label{ccd4}}
\end{center}
\end{figure}

The 3-point diagram shown in Fig.~\ref{ccd4} evaluates to:
\beq
\frac{-i m |g|^2 g^*}{2a}\int\frac{a^4 ~ d^4k}{(2\pi)^4}
\frac{ \nott{P}(p'+k) \( 1+ M(p+k) / s(p+k) \) }
  {r(p'+k) r(p+k) r(k) s(p'+k) } P_-
\eeq
Note that when the external momenta are set to $p=p'=0$,
this expression vanishes, due to an 
odd integrand, $\nott{P}(-k) = -\nott{P}(k)$.
Thus the Yukawa coupling receives no corrections in
the zero-momentum subtraction scheme.

The $Z-1$ counterterms (cf.~Eq.~\myref{react} below) are computed in the usual way, from the
sum of amputated one-particle irreducible self energy diagrams $\Sigma(p)$.  For instance in the case
of the scalar
\beq
\vev{\phit(p) \phit^*(p)} = \frac{1}{p^2+m^2} + \frac{1}{p^2+m^2} \Sigma(p) \frac{1}{p^2+m^2}
- \frac{1}{p^2+m^2} (Z_\phi-1) p^2 \frac{1}{p^2+m^2} + \cdots
\eeq
where $\cdots$ represents higher order terms.  Since there
is no mass renormalization at one loop we have $\Sigma(0)=0$ and
$\Sigma(p) \equiv |g|^2 \Sigma_2 ~ p^2 + \ord{p^4}$.
So, to have cancellation of $\Sigma_2$ at $p=0$ we require
$Z_\phi - 1 = |g|^2 \Sigma_2$.

\subsection{Numerical results}
We have computed the loop integrals in two ways:  working
with discrete sums at finite system size $L$, and
with numerical integration for infinite system size
$L \to \infty$.
We set the scale using the bare mass $m$.  In the
discrete sums, we use $mL=8$; the number
of lattice sites in any direction is $N=L/a=mL/ma$.
Equivalently, $ma = mL/N$ determines the lattice
spacing in units of $m$.  We have also computed with
$mL=16$ and find that the difference is only in the
third or fourth significant figure.  For the scalar we
compute $Z_\phi-1 = \lim_{p\to0} \Sigma_\phi(p)/p^2$, with $\Sigma_\phi(p)$ 
the one-loop self-energy Fig.~\ref{ccd1},
since $\Sigma(p)$ vanishes at $p=0$, as noted above.  The
results are given in Table \ref{sse}.  For the fermion we 
compute $Z_\chi-1= \lim_{p\to0} \Sigma_\chi(\nott{p})/\nott{p}$ 
since $\Sigma_\chi(\nott{p})$ vanishes
at $\nott{p}=0$. The results are given in Table \ref{fse}.
For the auxiliary field, we compute $Z_F-1 = \Sigma_F(0)$.
The results are given in Table \ref{ase}.

\begin{table}
\begin{center}
\begin{tabular}{|c|c|c|} \hline
$ma$ & $(Z-1)(\infty)/|g|^2$ & $(Z-1)(8)/|g|^2$ \\ \hline
0.5	& -0.00589 & -0.006355 \\
0.25	&  -0.00884 & -0.009532\\
0.125	& -0.01205 & -0.013358\\
0.0625	&-0.01573 & -0.017541\\
0.03125	&-0.01986 & -0.021837\\ 
0.015625&-0.02573 & -0.026205 \\
0.0078125 &-0.02956 &	-0.030583 \\ 
0.00390625 & -0.03370 & --- \\
0.001953125 & -0.03671 & --- \\ \hline
\end{tabular}
\caption{Self-energy for
the scalar, from Fig.~\ref{ccd1} evaluated at $p=(0,0,0,2\pi/L)$ 
for $L=\infty$ and $mL=8$.
\label{sse}}
\end{center}
\end{table}

\begin{table}
\begin{center}
\begin{tabular}{|c|c|c|} \hline
$ma$ & $(Z-1)(\infty)/|g|^2$ & $(Z-1)(8)/|g|^2$\\ \hline
0.5 &	-0.00385 & -0.004027 \\
0.25 &	-0.00715 &-0.006973 \\
0.125 &	-0.01085 &-0.011406 \\
0.0625 &	-0.01463 &-0.015632 \\
0.03125 &	-0.01893 &-0.019965 \\
0.015625 &	-0.02319 &-0.024336 \\
0.007813 &	-0.02910 & -0.028720 \\
0.00390625 & -0.03311 & --- \\
0.001953125 & -0.03697 & --- \\
\hline
\end{tabular}
\caption{One-loop counterterm $Z-1$ for
the fermion, evaluated from slope in self-energy
in $p \to 0$ limit, for $L=\infty$ and $mL=8$.  
Note that the lattice spacing is measured in units of $1/m$. \label{fse}}
\end{center}
\end{table}

\begin{table}
\begin{center}
\begin{tabular}{|c|c|c|} \hline
$ma$ & $(Z-1)(\infty)/|g|^2$ & $(Z-1)(8)/|g|^2$ \\ \hline
0.5 &	0.100785	&0.082542 \\
0.25 &	0.0994727&	0.088658 \\
0.125	&0.103388	&0.095266\\
0.0833333&	0.106955	&0.09916\\
0.0625	&0.10983	&0.105719\\
0.03125	&0.117483	&0.11255\\
0.015625&	0.125715	&0.13276\\
0.0078125&	0.134227	&0.140261\\
0.00390625&	0.14287365	&0.147476
\\ \hline
\end{tabular}
\caption{Self-energy for
the auxiliary field, evaluated at $p=0$ 
for $L=\infty$ and $mL=8$. \label{ase}}
\end{center}
\end{table}

We have fit the $L=\infty$ numerical data with $ma \leq 0.125$ to
\beq
f(ma) = c_0 \ln (ma) + c_1 ma + c_2 (ma)^2
\eeq
Giving the data points equal weight, the fit for the scalar self energy is
\beq
c_0= 0.00604(7), \quad c_1= 0.024(20), \quad c_2= -0.15(17)
\eeq
while for the fermion the fit is
\beq
c_0= 0.00597(5), \quad c_1= 0.061(15), \quad c_2= -0.39(12)
\eeq
Thus we see that at $L\to \infty$ the log divergences
of the scalar $Z_\phi-1$ and the fermion $Z_\chi-1$ match.
For the auxiliary field we obtain instead
\beq
c_0= -0.0261(5), \quad c_1= 0.87(11), \quad c_2= -3.9(1.0)
\eeq
so that its $Z_F-1$ does not match the scalar and fermion.
This is consistent with the results found in \cite{Kikukawa:2004dd}.

\subsection{Renormalization}
We can absorb all the logarithmic
divergence by rescaling the fields, so that the action is modified from \myref{lata}  and \myref{fmat} to
read
\beq
S &=& -a^4 \sum_x \bigg\{ \half Z_\chi \chit^T C M \chit - Z_\phi \phit^* D_2 \phit 
+ Z_F \Ftilb (1 - \frac{a}{2} D_2)^{-1}   \Ftil
\ddd + \sqrt{Z_F} \Ftilb (m^* \sqrt{Z_\phi} \phitb + g^* Z_\phi \phit^{* 2}) 
+ \sqrt{Z_F} \Ftil (m \sqrt{Z_\phi} \phit + g Z_\phi \phit^{ 2}) \bigg\},
\nnn M &=& \nott{D} + (m + 2 g \sqrt{Z_\phi} \phit) P_+ + (m^* + 2 g^* \sqrt{Z_\phi} \phit^*) P_-, 
\nnn \nott{D} &=& (1 - \half a D_2)^{-1} D_1
\label{react}
\eeq
It is interesting that when one eliminates $\Ftil,\Ftilb$ by their
equations of motion, the constant $Z_F$ disappears entirely:
\beq
S &=& -a^4 \sum_x \bigg\{ \half Z_\chi \chit^T C M \chit - Z_\phi \phit^* D_2 \phit 
\ddd - (m^* \sqrt{Z_\phi} \phitb + g^* Z_\phi \phit^{* 2}) 
(1 - \frac{a}{2} D_2) (m \sqrt{Z_\phi} \phit + g Z_\phi \phit^{ 2}) \bigg\},
\eeq
The discrepancy between $Z_F$ and $Z_\chi=Z_\phi$
is irrelevant to the physical theory, which only
contains $\phit$ and $\chit$.  In terms of the on-shell formulation,
the one-loop renormalization of the lattice theory,
in the $a \to 0$ limit, exactly mirrors what occurs in the continuum theory.
The mismatch $Z_F \not= Z_\chi = Z_\phi$ will have effects at 
two loops, where one-loop corrected
propagators will be involved in subdiagrams.  Cancellations
that in the continuum make use of equalities of counterterms
will no longer hold.  Thus we expect that nonsupersymmetric
renormalization of the on-shell formulation first appears
at two loops.

\subsection{Locality}
As stated above, we have computed $Z_\phi-1 = \lim_{p\to0} \Sigma_\phi(p)/p^2$.
The fact that this has a finite limit demonstrates
that at small $p$, the self-energy is analytic
in $p^2$.  Thus we find that
\beq
\Sigma_\phi(p) = \Sigma_2 p^2 + \ord{p^4}
\label{onse}
\eeq
and there is no evidence of nonlocality in the scalar self-energy.

\section{Supercurrent, mixing and renormalization}
\label{superc}
For a general superpotential $W(\phi)$, the supercurrent is
\beq
S^\mu = \sqtw \[ \nott{\p} \phi \gamma^\mu P_+ \chi
+ \nott{\p} \phi^* \gamma^\mu P_- \chi
+ \frac{\p W}{\p \phi} \gamma^\mu P_- \chi 
+ \(\frac{\p W}{\p \phi}\)^* \gamma^\mu P_+ \chi \]
\label{succ}
\eeq
and in our case $\p W/ \p \phi = m \phi + g \phi^2$.
Because of the supersymmetry breaking on the lattice,
this will mix with other operators in the same symmetry
channel.  For example, one has at the same naive dimension
the operator
\beq
T^\mu = \p^\mu \phi^* P_- \chi + \p^\mu \phi P_+ \chi
\eeq
If the lattice action (Eq.~\myref{mogeac} in the massive
case and Eq.~\myref{symgeac} in the massless case) is fine-tuned, then in the long
distance effective theory there will be a supercurrent
that is conserved in the continuum limit.  Thus
one way to detect \susy\ is to consider linear combinations
of bare lattice operators and extract the one
that has vanishing four-divergence in the supersymmetric
limit, modulo contact terms.  We will study that approach
in Section \ref{sufo} below.  Before doing so, we
briefly consider a naive discretization of the continuum
supercurrent \myref{succ}.

\subsection{Naive lattice supercurrent}
We formulate this lattice supercurrent in terms of tilded fields:
\beq
S^\mu &=& \sqtw \[ D_1 \phit \gamma^\mu P_+ \chit
+ D_1 \phit^* \gamma^\mu P_- \chit
+ \frac{\p W}{\p \phit} \gamma^\mu P_- \chit 
+ \(\frac{\p W}{\p \phit}\)^* \gamma^\mu P_+ \chit \] \nnn
\frac{\p W}{\p \phit} &=& m \phit + g \phit^2
\label{latsuc}
\eeq
For instance the correlation functions
\beq
\vev{S_\mu(x) \phit(y) \chit(0) C}, \quad \vev{S_\mu(x) \phit^*(y) \chit(0) C}
\label{sitre}
\eeq
give rise to tree-level diagrams from the quadratic terms in $S^\mu$.
It is straightforward to obtain at this order
\beq
\vev{S_\mu(x) \phit(y) \chit(0) C} = \sqtw \gamma_\nu \gamma_\mu P_- S(x) D_{1\nu}^{(x)} G(y-x)
+ \sqtw m^* \gamma_\mu P_+ S(x) G(y-x)
\eeq
where $S(x)= \vev{\chit(x) \chit(0) C}$ is the free theory 
fermion propagator and $G(x)=\vev{\phit(x) \phit^*(0)}$ is the free theory scalar propagator.
This will have the correct continuum limit $\p_\mu S_\mu=0$, modulo contact terms,
since no divergences appear at tree-level.  This can be explicitly
checked by differentiating the expression above and using the
equations satisfied by the free propagators:
\beq
&& D_1 S(x) = -(m P_+ + m^* P_-)S(x) + \ord{a}, \quad x \not= 0
\ddd D_1 D_1 G(x) = |m|^2 G(x) + \ord{a}, \quad x \not= 0
\eeq

The variation of the action under the lattice
version of the supersymmetry transformation \myref{stuntil} and \myref{staux} is
\beq
\delta S = \sqtw a^4 \sum_x \chit^T C \[g P_+ (2 \phit D_1 \phit - D_1 \phit^2) 
+ g^* P_- (2 \phit^* D_1 \phit^* - D_1 \phit^{*2}) \] \e
\label{niac}
\eeq
Note that this is $\ord{a}$, since
\beq
\lim_{a \to 0} \sum_x \chit^T C P_+ (2 \phit D_1 \phit - D_1 \phit^2) = 0
\eeq
Also note the presence of $g$ in \myref{niac}.
In order to see violation of the supersymmetric identity $\p_\mu S_\mu=0$
in the continuum limit, diagrams involving the coupling $g$ must be included.
Loop diagrams are required, in order to get the diverenges that
cancel the $\ord{a}$ factor coming from \myref{niac} in the continuum limit.

\subsection{Exactly conserved supercurrent?}
Because of \myref{niac}, one can ask whether 
in the $g=0$ limit there is an exactly conserved supercurrent;
by a proper latticization of the continuum
supercurrent and the four-divergence, we would hope to find 
a discretized version of $\p_\mu S_\mu=0$ that holds at finite lattice spacing.
In order to search for this hypothetical supercurrent, we follow the
usual method and suppose that the parameter of the
transformation $\e$ depends on spacetime position, $\e=\e(x)$.
Variation of the action with site-dependent $\e$ can generally
be written in the form:
\beq
\Delta S = -a^4 \sum_x \e^T(x) C {\mathfrak{S}}(x)
\label{uity}
\eeq
Thus in the form of the action with all fields, Eq.~\myref{kisu},
\beq
{\mathfrak{S}} &=& - \frac{1}{a^4} \frac{\delta S}{\delta (\e^T C)} 
\nnn
&=& - \sqtw \big( D_1 \phi D_1 P_- \chi + D_1^2 \phi P_- \chi
+ D_1 \phi D_2 P_+ \chi + D_2 \phi D_1 P_+ \chi
\ddd
+ D_1 \phi^* D_1 P_+ \chi + D_1^2 \phi^* P_+ \chi
+ D_1 \phi^* D_2 P_- \chi + D_2 \phi^* D_1 P_- \chi
\ddd
+ D_2 F P_+ \chi - F D_2 P_+ \chi
+ D_2 F^* P_- \chi - F^* D_2 P_- \chi \big)
\ddd
- \sqtw m \big( D_1 \phit P_+ \chit + \phit D_1 P_+ \chit \big)
- \sqtw m^* \big( D_1 \phit^* P_- \chit + \phit^* D_1 P_- \chit \big)
\ddd
- \sqtw g \big( 2 \phit D_1 \phit P_+ \chit + \phit^2 D_1 P_+ \chit \big)
- \sqtw g^* \big( 2 \phit^* D_1 \phit^* P_- \chit + \phit^{*2} D_1 P_- \chit \big)
\ddd
+ \frac{2}{a} \sqtw \big( D_1 \Phi P_+ X + \Phi D_1 P_+ X + D_1 \Phi^* P_- X + \Phi^* D_1 P_- X \big)
\eeq
On the other hand, when we just work with tilded fields, Eq.~\myref{lata},
\beq
{\mathfrak{S}} &=& -\sqtw ( D_1 \phit \nott{D} P_- \chit - \Ftil \nott{D} P_- \chit
+ D_1 \phit^* \nott{D} P_+ \chi - \Ftil^* \nott{D} P_+ \chit )
\ddd
+  \sqtw ( \frac{2}{a} D_2 \phit P_- \chit + \frac{2}{a} D_2 \phit^* P_+ \chit )
\ddd
- \sqtw ( D_1 P_- \chit (1- \frac{a}{2} D_2)^{-1} \Ftil
+ D_1 P_+ \chit (1- \frac{a}{2} D_2)^{-1} \Ftil^* )
\ddd
- \sqtw m \big( D_1 \phit P_+ \chit + \phit D_1 P_+ \chit \big)
- \sqtw m^* \big( D_1 \phit^* P_- \chit + \phit^* D_1 P_- \chit \big)
\ddd
- \sqtw g \big( 2 \phit D_1 \phit P_+ \chit + \phit^2 D_1 P_+ \chit \big)
- \sqtw g^* \big( 2 \phit^* D_1 \phit^* P_- \chit + \phit^{*2} D_1 P_- \chit \big)
\eeq
Finally, for the quadratic terms there exist lattice identities analogous to integration by parts,
such as:
\beq
\sum_x D_1 \phi D_1 P_- \chi = -\sum_x \phi D_1^2 P_- \chi, \quad
\sum_x D_2 \phi D_1 P_+ \chi = \sum_x \phi D_1 D_2 P_+ \chi
\eeq
These sorts of identities give us, for either form of $\mathfrak{S}$,
\beq
\sum_x a^4 \mathfrak{S} & = & - \sqtw \sum_x a^4 \bigg\{
 g \big( 2 \phit D_1 \phit P_+ \chit + \phit^2 D_1 P_+ \chit \big) \nnn
&& g^* \big( 2 \phit^* D_1 \phit^* P_- \chit + \phit^{*2} D_1 P_- \chit \big)
\bigg\}
\label{nzav}
\eeq
In the continuum limit this quantity also vanishes; however, on the
lattice we cannot integrate this cubic quantity by parts and
the expression inside braces gives an $\ord{a}$ deviation from zero.

In conclusion we do not find an exactly conserved supercurrent at $g=0$
but we do find an expression for which $\sum_x \mathfrak{S} = 0$,
behaving as if $\mathfrak{S} \sim \p_\mu S_\mu$, for $g=0$.  The
quantity $\mathfrak{S}$ will play an important role in the
broken \susy\ Ward-Takahashi identities of the lattice theory,
which we consider next.


\subsection{Broken Ward-Takahashi identities}
As we have just seen, the variation of the action under an $x$-dependent
spinor parameter $\e_\alpha(x)$ cannot be written as a simple product
of $\e$ and a finite difference operator acting on an expression like \myref{latsuc}.
Otherwise \myref{nzav} would vanish.
Rather, it takes a more general form, which we have denoted above as \myref{uity}.
Thus $\mathfrak{S}$ will appear in the broken Ward-Takahashi identities
that we are about the derive.
Parenthetically, in the classical continuum limit, with the discretization \myref{latsuc} of the supercurrent,
\beq
\lim_{a \to 0} ( {\mathfrak{S}}_\alpha(x) - \p_\mu S_{\mu,\alpha}(x) ) = 0
\eeq
because we know that each term should go over to the continuum
expression $\p_\mu S_{\mu,\alpha}(x)$.  

In addition to \myref{uity}, we also must consider the variation of the source terms, 
\beq
S_{src} = -a^4 \sum_x \bigg\{ {\tilde\eta}^T C P_+ \chit + {\tilde\tau}^T C P_- \chit
+ {\tilde J}^* \phit + {\tilde J} \phit^* + {\tilde k}^* \Ftil + \tilde k \Ftil^* \bigg\}
\label{Ssrct}
\eeq
For this purpose we define $\delta_\e \phit = (\e^T C)_\alpha \Delta_\alpha \phit$,
etc.  The \susy\ transformations of the tilded fields are given in \myref{sttil}.  Thus
\beq
&&
\Delta_\alpha \chit_\beta = -\sqtw ( P_+ (D_1 \phit + \Ftil) C )_{\beta \alpha}
- \sqtw (P_- (D_1 \phit^* + \Ftil^*) C)_{\beta \alpha},
\ddd
\Delta_\alpha \phit = \sqtw ( P_+ \chit)_\alpha, \quad
\Delta_\alpha \phit^* = \sqtw ( P_- \chit)_\alpha
\ddd \Delta_\alpha \Ftil = \sqtw ( D_1 P_+ \chit)_\alpha, \quad 
\Delta_\alpha \Ftil^* = \sqtw ( D_1 P_- \chit)_\alpha
\eeq
Then we have the identity
\beq
&& \Delta_\alpha Z( {\cal J}; x) = 0
= \int [ d \phit d \phit^* d \chit d \Ftil d \Ftil^*] e^{-S - S_{src}}
\bigg[ 
{\mathfrak{S}}_\alpha(x) 
\ddd
+ ({\tilde \eta}^T C P_+ \Delta_\alpha \chit)(x)
+ ({\tilde \tau}^T C P_- \Delta_\alpha \chit)(x)
\ddd
+ ({\tilde J}^* \Delta_\alpha \phit)(x)
+ ({\tilde J} \Delta_\alpha \phit^*)(x)
+ ({\tilde k}^* \Delta_\alpha \Ftil)(x)
+ ({\tilde k} \Delta_\alpha \Ftil^*)(x) \bigg]
\label{luytr}
\eeq
Here ${\cal J}$ collectively denotes all the sources and $S$ is the
action \myref{lata}.
Since the identity holds for all values of ${\cal J}$,
derivatives of $\Delta Z( {\cal J}; x)$ with
respect to the sources also vanish.  Thus it is the
generating function of lattice correlation function identities
associated with the \susy\ transformations \myref{sttil}.
We refer to these as the broken \susy\ Ward-Takahashi identities.

We could measure a lattice transcription of the Ward-Takahashi identities
that vanish in the continuum theory.  Their deviation
from zero gives a measure of the amount of \susy\ violation
in the latticized theory.  A simple example is the following.
\beq
&& \( \frac{ \delta }{\delta (\tilde\eta^T C)_\beta(y)} + 
\frac{ \delta }{\delta (\tilde\tau^T C)_\beta(y)} \) \Delta_\alpha Z ({\cal J},x) \bigg|_{{\cal J}=0}
= 0
\nnn
&& = \vev{ - a^4 {\mathfrak{S}}_\alpha(x) \chit_\beta(y) - \sqtw (( D_1 \phit + \Ftil) C )_{\beta \alpha}(x) \delta(x,y) }
\eeq
This leads us to the identity
\beq
\sqtw \sum_x a^4 \vev{ \Ftil(x) } C_{\alpha \beta} = -\sum_{x,y} a^8 \vev{ {\mathfrak{S}}_\alpha(x) \chit_\beta(y) }
\label{exid}
\eeq
In the continuum theory the right-hand side vanishes because $\mathfrak{S} \to \p_\mu S_\mu$.
On the lattice we have instead the simplified expression \myref{nzav},
which is an $\ord{a}$ lattice artifact.  Divergent behavior in $\vev{ {\mathfrak{S}}_\alpha(x) \chit_\beta(y) }$
can lead to a result which does not vanish in the continuum limit.
We can measure the violation of the continuum \susy\ Ward-Takahashi identity
$\int d^4 x \vev{F(x)}=0$ by computing either side of \myref{exid} in our numerical simulations.
This is done for the left-hand side in Section \ref{wtis} below.

Consider again the most general lattice action consistent with symmetries, Eq.~\myref{mogeac}.
By tuning the parameters in this bare action we expect to obtain the supersymmetric limit
for the long distance effective theory.  
On the other hand, once this fine-tuning is performed the bare lattice
action \myref{mogeac} will not be invariant with respect to the \susy\ transformation of the
bare fields:
\beq
\delta S = - a^4 \e^T C \sum_x \mathfrak{S} \not= 0
\eeq
Thus we continue to have modified identities, \myref{luytr}, which
do not look qualitatively different from those away from the supersymmetric
point in parameter space.  What is different is that in the long
distance effective theory there is a conserved supercurrent,
$\p_\mu S_\mu=0$ as an operator relation in the continuum limit.  To probe for \susy, 
we must construct this supercurrent, built out of the appropriate
set of bare field operators, Eq.~\myref{fpas} below.

\subsection{Supercurrent formulation}
\label{sufo}
We write down all operators that have the index
structure of $S_{\mu \alpha}(x)$; leading representatives are
given in Table \ref{genops}.  We denote these
as $\Ocal_{\mu \alpha, j}^{(n/2)}$ where $n/2$ is the engineering dimension.
A linear combination of these is the long distance effective supercurrent 
at lattice spacing $a$:
\beq
S_{\mu \alpha}(x) = \sum_{n=3,5,7,\ldots} \sum_j b_j^{(n/2)}(a) a^{(n-7)/2} \Ocal_{\mu \alpha, j}^{(n/2)}(x)
\label{fpas}
\eeq
On dimensional grounds, $b_j^{(n/2)}(a)$ must be a dimensionless quantity,
and is therefore a function of the renormalized parameters of the
long distance theory, $m_r a$ and $g_r$, in the infinite volume limit.
At finite volume it may also depend on $a/L$.  We know that
\beq
\lim_{a \to 0} \p_\mu S_{\mu \alpha}(x) = 0
\label{scc}
\eeq
as an operator relation, provided the relevant and marginal counterterms
in the lattice action \myref{mogeac} are tuned properly.  If the fermion
mass is set to zero, then we only need to tune the lattice action \myref{symgeac}.
In practice, we will add a fermion mass term to \myref{symgeac}
and extrapolate to vanishing bare fermion mass.

In numerical computations it is convenient to take the spatial transform
so that we instead work with 
\beq
a^3 \sum_{\bm x} ~ \p_\mu S_{\mu \alpha}(t, \bm x) = a^3 \sum_{\bm x} ~ \p_t S_{0 \alpha}(t, \bm x)
= \p_t Q_\alpha(t) .
\eeq
Since $S_{\mu \alpha}$ is fermionic, an odd number of $\chit$ fields must appear
in nonvanishing correlation functions with $\p_t Q_\alpha(t)$.
For this reason, correlation functions should involve operators with dimension
of at least 3/2.  In what follows we will exclusively consider
two-point functions with fermionic operators of dimension $n/2$ which we denote $\Ocal_j^{(n/2)}$,
where $n=3,5,7$ etc.  In order to identify the supercurrent nonperturbatively,
what one really wants to study is the family of correlation functions
\beq
M_{ij}^{(m,n)}(t) = \vev{ \p_t \Ocal_{0 \alpha, j}^{(n/2)} (t) \Ocal_i^{(m/2)}(0, \bm 0) }
\label{gencor}
\eeq
where
\beq
\Ocal_{0 \alpha, j}^{(n/2)} (t) = a^3 \sum_{\bm x} ~ \Ocal_{0 \alpha, j}^{(n/2)} (t, \bm x)
\eeq
Truncating $n,m \leq N_{\text{max}}$, we want to tune the
parameters of the bare action \myref{symgeac} such that $M_{ij}^{(m,n)}(t)$
has a null space in the collective column index $B=(j,n)$ for each $i, m$ 
and $t$, which together form a collective row index $A=(i,m,t)$:
\beq
\lim_{a \to 0} M_{AB} b_B=0
\label{matprob}
\eeq
for all $A=1,\ldots,N_c$, with $b$ nontrivial.

Of course at finite lattice spacing, no amount of fine-tuning
will restore supersymmetry.  Thus what we seek is not \myref{matprob},
but instead $M_{AB} b_B= \ord{a}$.  However we need to find
$b_B$ with the constraint $||b||=1$ in order to make the tuning
independent of the normalization of this vector.  We thus seek
to minimize
\beq
F = \sum_{A} \big| \sum_B M_{AB} b_B \big|^2, \quad \sum_B |b_B|^2 \equiv 1
\eeq
with respect to $b$.  This is repeated at various $m_2^2$ and $\lambda_1$
until the minimum is found with respect to these parameters for fixed
$m_1$.  Finally, we make an extrapolation to $m_1=0$, and identify the
fine-tuned pair $m_2^2$ and $\lambda_1$ where $F$ approaches its minimum value.

We have taken $||b||=1$ in our considerations so far.
But we know that is not the whole story in the $a \to 0$ limit.
The operator $S_\mu$ will undoubtedly need 
to be renormalized, and that will involve a divergent factor:  $S_\mu^{ren.} = Z_S S_\mu$.  
This can be determined through a position space scheme.
We demand that the free theory
results are matched at some distance scale $r$:
\beq
Z_S^2 \vev{S_\mu(x) S_\nu(0)}_{|x|=r} = \vev{S_\mu(x) S_\nu(0)}_{\text{free},|x|=r}
\eeq
Similarly, for the operators that appear in  the correlation
functions with $S_\mu$ we must impose
\beq
(Z_i^{(m/2)})^2 \vev{\Ocal_i^{(m/2)}(x) \Ocal_i^{(m/2)}(0)}_{|x|=r} 
= \vev{\Ocal_i^{(m/2)}(x) \Ocal_i^{(m/2)}(0)}_{\text{free},|x|=r}
\eeq
Then the quantity that should be minimized, and which must vanish in
the continuum limit is:
\beq
F = \sum_{t,m,i} \big| Z_S Z_i^{(m/2)} \sum_{j,n} b_j^{(n/2)} \vev{
\p_t \Ocal_{0\alpha,j}^{(n/2)}(t) \Ocal_i^{(m/2)}(0, \bm 0)} \big|^2
\eeq

\begin{table}
\begin{center}
\begin{tabular}{|c|c|} \hline
dimension & operators \\ \hline
3/2 & $\gamma_\mu P_\pm \chi$ \\ \hline
5/2 & $\gamma_\mu \phi P_\pm \chi$, $\gamma_\mu \phi^* P_\pm \chi$,
$P_\pm \p_\mu \chi$, $\s_{\mu \nu} P_\pm \p_\nu \chi$ \\ \hline
7/2 & $\gamma_\mu \phi^2 P_\pm \chi$, $\gamma_\mu \phi^{*2} P_\pm \chi$, 
$\gamma_\mu |\phi|^2 P_\pm \chi$, $\p_\mu \phi P_\pm \chi$, $\p_\mu \phi^* P_\pm \chi$,
$\phi P_\pm \p_\mu \chi$, $\phi^* P_\pm \p_\mu \chi$, $\nott{\p} P_\pm \p_\mu \chi$, \\
    &
$\s_{\mu \nu} \p_\nu \phi P_\pm \chi$, $\s_{\mu \nu} \p_\nu \phi^* P_\pm \chi$,
$\phi \s_{\mu \nu} \p_\nu P_\pm \chi$, $\phi^* \s_{\mu \nu} \p_\nu P_\pm \chi$
\\ \hline
\end{tabular}
\caption{Operators that can potentially mix with the supercurrent,
up to $\ord{a}$ supressed higher dimensional operators. \label{genops}}
\end{center}
\end{table}

\subsection{Continuum limit}
Since the model is a $|\phi|^4$ theory coupled to fermions through a Yukawa interaction,
it is expected to be trivial.  The only continuum limit is therefore the free theory.
We can work at arbitrarily small but finite lattice spacing $a$.  Having tuned to the
supersymmetric point in parameter space there is only one renormalized mass parameter $m_r$
in the theory, and so in lattice units we will measure $m_r a$ from the
exponential decay of correlation functions of the elementary fields.
Due to the $U(1)_R$ symmetry, this mass will be proportional to the
bare fermion mass, which in lattice units is $m_1 a$.  
It follows that we take $a$ smaller by reducing the
size of the lattice parameter $m_1 a$.  The triviality of the
theory is then the statement that if $m_1 a=0$, the long
distance effective coupling $g_r=0$, independently of the
bare coupling $y_1$ (recall that $\lambda_1$ is fine-tuned
to obtain the supersymmetric limit).

\section{Fine-tuning with multicanonical reweighting}
\label{mcrw}
Multicanonical methods \cite{Binder,Baumann:1986iq,Berg:1991cf}
combined with ``Ferrenberg-Swendsen 
reweighting'' \cite{Falcioni:1982cz,Ferrenberg:1988yz,Ferrenberg:1989ui}
have proven
to be a powerful tool for maximizing the usefulness
of Monte Carlo simulations over a range of parameter space.
We refer to this combination of techniques as multicanonical reweighting (MCRW).
For instance, MCRW was applied in a study comparing $SU(2)$ and $SO(3)=SU(2)/Z_2$
lattice gauge theories \cite{deForcrand:2002vx,deForcrand:2002vs}.  It
was found to dramatically flatten the distributions
with respect to three parameters, twists on gauge fields
at the spatial boundaries.  Another successful
application of MCRW 
consists of lattice results for the electroweak
phase transition \cite{Kajantie:1995kf,Moore:2000jw}.

We will begin by describing  
MCRW generally for a theory of a real scalar field $\phi$, followed by a presentation
of how it would be applied to the lattice four-dimensional
Wess-Zumino model that we are studying.

\subsection{Preliminaries}
Suppose we perform a Monte Carlo simulation at one value $m_0$ of the scalar mass $m$,
so that the configurations sample the distribution determined
by the action 
\beq
S(m = m_0) = S(m=0) + \half \int d^4 x ~ m_0^2 \phi^2(x) .
\label{m0act}
\eeq
Following the ``Ferrenberg-Swendsen reweighting'' method 
\cite{Falcioni:1982cz,Ferrenberg:1988yz,Ferrenberg:1989ui}
one can use the following
{\em reweighting identity} to compute the expectation value $\vev{ \Ocal }_m $ of an operator
$\Ocal$ for the distribution with a mass $m$:
\beq
\vev{ \Ocal }_m = \frac{ \vev{\Ocal e^{- \Delta S(m)} }_{m_0} }{ \vev{ e^{- \Delta S(m)} }_{m_0} }
\approx \frac{\sum_{C \in F(n)} 
\Ocal_C  \exp \[ -\half ({m}^2-m_0^2) \int d^4 x ~ \phi^2_C \] }
{\sum_{C \in F(n)} \exp \[ -\half ({m}^2-m_0^2) \int d^4 x ~ \phi^2_C \] } .
\label{ilte}
\eeq
In the first equality $\vev{ \cdots }_{m_0}$ is the expectation value with
respect to the canonical distribution corresponding to \myref{m0act} and
\beq
\Delta S(m) = \half ( m^2-m_0^2) \int d^4  x ~ \phi^2
\eeq
is the shift in the action when the mass is changed.  In the second,
$\int d^4x ~ \phi^2_C$ and $\Ocal_C$ are the mass term
operator evaluated on configuration $C$
and $\sum_{C \in F(n)}$ is the sum over the distribution
$F(n)$ of $n$ configurations generated in the Monte
Carlo simulation.  These of course provide a finite
ensemble that approximates the canonical distribution
corresponding to \myref{m0act}.  The advantage of this
approach is that one need only perform a single simulation
at mass $m_0$, storing the values of $\int d^4x ~ \phi^2_C$ and 
$\Ocal_C$ for each $C$, and then $\vev{ \Ocal }_m$ can
be computed for a swath of the parameter space $m$ without
having to perform any new simulations.  Typically the
time for this ``offline'' calculation is negligible compared
to that of the simulation.

Unfortunately, the regime of utility for this technique is limited
by the {\it overlap problem,} in a way that often worsens
exponentially in the spacetime volume.  For instance,
suppose the theory \myref{m0act} has a quartic interaction
and a critical mass-squared $m_c^2$ such that for $m^2 < m_c^2$ there
is spontaneous symmetry breaking.  If we simulate with
$m_0^2 > m_c^2$ then the field is exponentially weighted
toward $\int d^4x ~ \phi^2 \approx 0$.  Now suppose we
attempt to reweight to $m^2 < m_c^2$.  In that case $-(m^2-m_0^2)>0$
so that the exponential weight factor in \myref{ilte} is
minimal at $\int d^4x ~ \phi^2 \approx 0$.  The ensemble
that is generated in the Monte Carlo simulation will have
exponentially few configuration in the regime where
$\int d^4x ~ \phi^2$ is far from zero and $e^{-\Delta S(m)}$
is large.  Because we will have very few representatives
of configurations with the largest weight $e^{-\Delta S(m)}$,
and most members of the ensemble have very small weight, fluctuations
will be large and huge samples are 
required in order to have acceptable errors.  
The mismatch of the distributions gets worse as 
the number of lattice sites increases, because
the exponent is extensive (i.e., scales like the spacetime
volume $L^3 \times T$).  

As a concrete example, Fig.~4 of \cite{deForcrand:2002vs} shows that in
the range of a three-dimensional parameter space the
ordinary canonical Monte Carlo density of states varies by 14
orders of magnitude.  This is for an $8^3 \times 4$ lattice,
which is still relatively small.  The problem will get exponentially
worse on larger lattices.

In a number of contexts 
the technique of {\it multicanonical reweighting} \cite{Binder,
Baumann:1986iq,Berg:1991cf}
has been found to ameliorate the overlap problem.
One replaces $S$ with  
\beq
S_{MCRW}=S+W[\Ocal_1, \Ocal_2, \ldots], 
\label{srww}
\eeq
where $W[\Ocal_1, \Ocal_2, \ldots]$ is a
carefully engineered function of some small set of observables.
For instance in the model that we are studying,
$W$ will be a function of the mass term and the
quartic term.
\beq
\Ocal_1 = a^4 \sum_x ~ |\phi|^2, \quad \Ocal_2 = a^4 \sum_x ~ |\phi|^4.
\eeq
The (reweighted) expectation value of an observable in the distribution corresponding
to $S_{MCRW}$ is:
\beq
\vev{\cal O} = \frac{\sum_{C \in F(n)} \Ocal_C \; e^{ W[\Ocal^C_1, \Ocal^C_2] }}
{\sum_{C \in F(n)} e^{ W[\Ocal^C_1, \Ocal^C_2]} } .
\label{rwww}
\eeq

Since the $e^W$ factor in \myref{rwww} just cancels the $e^{-W}$ Boltzmann factor
coming from \myref{srww}, one might wonder why it is introduced in the first place.
The point is that the additional Boltzmann factor $e^{-W}$ in effect
produces a weighted average over a continuum of canonical ensembles 
(hence the appelation ``multicanonical'') such that there is
a good overlap with the distribution that one is
reweighting to.  The challenge is to design a $W$ such that
sampling is flattened over the range of observables one is interested in.

We return to Fig.~4 of \cite{deForcrand:2002vs}.
It shows that the multicanonical Monte Carlo sampling distribution is flat in
the range of three-dimensional parameter space between the
peaks, where the
ordinary canonical Monte Carlo distribution varies by twelve
orders of magnitude.  The reweighting function $W$ was
represented by a numerical table, composed of the inverse
density of states with respect to the tuned parameters.
This is for an $8^3 \times 4$ lattice,
which is still relatively small, and it indicates
that $\ord{10^{12}}$ more samples would be required
in the canonical Monte Carlo approach in order
to scan a comparable range of parameter space
by ordinary Ferrenberg-Swendsen reweighting techniques.

As another example, in studying first order phase transitions 
(e.g., \cite{Kajantie:1995kf}), one chooses
$\Ocal_1$ to be the order parameter of the transition; in a model with a
scalar field, typically $\Ocal_1 = \int d^4 x ~ \phi^2$.  One tunes $W[\Ocal_1]$ to
cancel the nonperturbative effective potential for this operator, so that
the Monte Carlo simulation samples evenly in $\Ocal_1$.  This enhances
statistics for configurations intermediate between the phases.
In the mass scan example of Eq.~\myref{ilte}, one has
\beq
\vev{\Ocal} = \frac{\sum_{C \in F(n)} \Ocal_C \exp \( W[\int d^4 x ~ \phi_C^2] 
- \half (m_2^2-m_1^2) \int d^4 x ~ \phi^2_C \) }
{\sum_{C \in F(n)} 
\exp \( W[\int d^4 x ~ \phi_C^2] -\half (m_2^2-m_1^2) \int d^4 x ~ \phi^2_C \) } .
\label{omex}
\eeq
In this way, wherever the exponential in \myref{omex} 
happens to be at its maximum,
a large number of configurations will be generated, due to
the flat distribution with respect to $\int d^4 x ~ \phi^2$.

Two approaches exist for engineering a good function $W$.
\bit
\item[(1)]
One can employ a bootstrap method that iterates between Monte Carlo 
simulation and adjustments to $W$.  For
instance a numerical tabulation of density of
states $\rho$ may be obtained from a canonical simulation,
as was done in \cite{deForcrand:2002vx,deForcrand:2002vs}.  Schematically,
one obtains a histogram estimate
of $\rho(\Ocal_1)$ for an operator value range $\Ocal_1$ range $\Ocal_{1,\text{min}}
\leq \Ocal_1 \leq \Ocal_{1,\text{max}}$.  This provides
an initial version of $W$, through $W \equiv 1/\rho(\Ocal_1)$.
If necessary, the process can be repeated to refine the table.
\item[(2)]
Iterative or stochastic searches may be used to
optimize $W$ with respect to a predetermined parameterization in a small volume.
Performing this at two different small volumes then provides
an extrapolation estimate for $W$ in the next largest volume,
which can then be refined through another search.
\eit

\subsection{Application}
If we work at $m_1 \not= 0$,
the action that we must tune nonperturbatively is \myref{mogeac}.
If extrapolate to the $m_1 \to 0$ limit, $U(1)_R$
symmetry \myref{tiltra} can be imposed and we must fine-tune
\myref{symgeac}.  We fix $y_1$, so that it determines the coupling strength that we
study.  All other parameters, $m_2^2$ and $\lambda_1$, are associated with bosonic terms
in the action, and can be tuned offline in the way that was just
described above.  

\section{Simulation}
\subsection{Pfaffian phase}
The integration over lattice fermions yields
\beq
{\rm Pf} (C M), \quad M = \nott{D} +  m P_+ + m^* P_- + g \phit P_+ + g^* \phitb P_-
\eeq
The transformation
\beq
\chit(x^0,\bm x) \to \gamma^0 \chit(x^0, - \bm x), \quad
\phit(x^0,\bm x) \to \phit^*(x^0, - \bm x), \quad
m \leftrightarrow m^*, \quad
g \leftrightarrow g^*
\eeq
leaves $\chit^T C M \chit$ invariant.
Its effect on the Pfaffian is then
\beq
{\rm Pf} (C M)(m,g,\phit) = {\rm Pf} (C M)(m^*,g^*,\phit^*) = [{\rm Pf} (C M)(m,g,\phit)]^*
\eeq
so that the Pfaffian is real.  There is still the possibility of a sign problem.
At weak coupling and $|m|$ not too small this is likely circumvented, since
the spectrum of $M$ is pushed away from zero and crossings
are avoided.  In that case
we can work with the phase quenched measure,
\beq
|{\rm Pf} (C M)| = \det (M^\dagger M)^{1/4}
\eeq

\subsection{Pseudofermions}
The Pfaffian is represented through an integration over bosonic fields, the pseudofermions $\eta$, with action
\beq
S_{PF} = \eta^\dagger (M^\dagger M)^{-1/4} \eta .
\eeq
We compute this using the rational approximation
\beq
(M^\dagger M)^{-1/4} \eta \approx \alpha_0 \eta + \sum_{i=1}^d 
\frac{\alpha_i}{M^\dagger M + \beta_i} \eta
\label{jkjk}
\eeq
where the coefficients $\alpha_i, \beta_i$ are chosen
to minimize errors over the range of eigenvalues of
the operator $M^\dagger M$.  Since we will work at weak
coupling in what follows, we are able to use the spectrum
of the $g=0$ operator, $M_0^\dagger M_0$, to determine this range.
In \myref{jkjk}, the integer $d$ is the degree of the approximation.
Evaluation of \myref{jkjk} requires us to solve the linear algebra problem
\beq
(M^\dagger M + \beta_i ) X_i = \eta
\eeq
for each degree $i=1,\ldots,d$.
This is done for each $i$ by conjugate gradient, which is an iterative
solver designed for sparse linear systems.  In computing the $M^\dagger M$
matrix multiplication, we encounter $M f_{in} = f_{out}$ and
a similar expression involving $M^\dagger$.
There is a trick for the computation of the ``dslash'' $\nott{D}$; we
fast Fourier tranform the in vector $f_{in}$ to momentum
space, apply dslash here where it is diagonal (giving
$f_{out}$ in momentum space), then fast Fourier transform
back to position space.  This avoids an additional layer of rational
approximation for computing products involving $(A^\dagger A)^{1/2}$
in position space.

\begin{table}
\begin{center}
\begin{tabular}{|c|c|c|c|c|c|c|} \hline
Lattice & CPU & GPU (CUDA) & GPU (Ours) & CPU  & GPU (CUDA) & GPU (Ours) \\ 
        & single & single & single & double & double & double \\ \hline
$8^3 \times 32$ & 1.1 & 6.9 & 24 & 0.94 & 4.4 & 11 \\
$16^3 \times 32$ & 0.88 & 14 & 71 & 0.69 & 10 & 35 \\
$32^3 \times 64$ & 0.11 & 20 & --- & 0.085 & 10 & --- \\
\hline
\end{tabular}
\caption{Comparison of timing, Gflop/s, for fast fourier transform of
the spinor field.  Both single and double precision results are given. \label{tabfftt}}
\end{center}
\end{table}

In Table \ref{tabfftt} we give a comparison of timing for fast fourier transform (FFT) of
a lattice spinor field such as $f_{in}$ using different tools.  Concerning hardware,
the CPU runs are from an Intel Xeon Woodcrest 5160, whereas the GPU runs are
from a Nvidia GTX 285.  The CPU runs were performed using the four-dimensional
Numerical Recipes code \cite{NR}.  The first version of GPU code performs four one-dimensional
transforms using the batched CUDA FFT available from Nvidia to get the four-dimensional transform.  The
second version of GPU code uses our own FFT for the spatial dimensions, which is currently only
operational for $L < 32$, due to register constraints.  As can be seen,
it is quite a bit faster than CUDA FFT.  In the temporal dimension $T \geq 32$
so the CUDA FFT is always used in that direction.
A necessary reorganizing of the arrays after each of the four one-dimensional
transforms is included in the timing.  We do not know the
origin of the drop in performance of the Numerical Recipes code on large lattices,
though it may be related to non-cached memory accesses.  
We inserted a counter of floating point operations (flop) into the
Numerical Recipes code and found that the number of operations
was approximately given by $5 N \log_2 N$.
We used the estimate $5 N \log_2 N$ for the 
number of operations for GPU code as well.  What one sees is
that even on small lattices the FFT runs an order of
magnitude faster using the GPU.  Since this is
the crucial operation in applying the lattice Dirac
operator, we anticipate simulation speeds of approximately 10 Gflop/s
for the $16^3 \times 32$ and larger lattices, working
in double precision.  In what follows we use our own FFT for
the $L < 32$ lattices, but the CUDA FFT for the $L=32$ lattice.

\subsection{Preconditioning}
At weak couplings, we expect that preconditioning by the inverse of the
free theory fermion matrix $M_0$ will improve convergence of the
conjugate gradient solver.  We must formulate the problem in terms
of a hermitian, positive definite matrix.  For this purpose,
we re-express the problem 
\beq
M^\dagger M x = b
\label{theprob}
\eeq
as follows:
\beq
(M_0^{-1 \dagger} M^\dagger) (M M_0^{-1}) (M_0 x) = (M_0^{-1 \dagger} b)
\eeq
This leads to the definitions
\beq
\tilde M = M M_0^{-1}, \quad \tilde x = M_0 x, \quad \tilde b = M_0^{-1 \dagger} b
\eeq
The problem in terms of these variables is $\tilde M^\dagger \tilde M \tilde x = \tilde b$.
Once we obtain $\tilde x$ from the conjugate gradient solver, we obtain the
desired solution from $x = M_0^{-1} \tilde x$, where $M_0^{-1}$ can be computed
analytically.  It is just
\beq
M_0^{-1} = \frac{M_0^\dagger}{D_\mu^\dagger D_\mu + |m|^2}, \quad
\nott{D} \equiv \gamma_\mu D_\mu = (1 - \frac{a}{2} D_2)^{-1} D_1
\eeq

We have solved the problem $M^\dagger M x = b$ with random Gaussian
$b$ on lattices of various sizes, using our GPU code.  Convergence is
obtained with and without preconditioning.  A comparison is given in
Table \ref{invtab1}.  The speed-up factor from the preconditioning
is quite large, and the inversion times are rapid enough that simulations
on relatively large lattices are realistic.

\begin{table}
\begin{tabular}{|c|c|c|c|c|c|c|c|c|}
\hline
Lattice        & Precision  & NPC secs. & NPC iters. & Gflop/s & PC secs. & PC iters. & Gflop/s & speed-up \\
\hline
$8^3 \times 32$ & single    &  1.3      &    830     & 14       &  0.038  &   8     &  17     & 34  \\
$8^3 \times 32$ & double    &  4.9      &   1600     & 7.2      &  0.13   &   15    &  8.1    & 38  \\
$16^3 \times 32$ & single   &  7.1      &    870     & 25       &  0.20   &   8     &  31     & 36  \\
$16^3 \times 32$ & double   &  31       &   1800     & 12       &  0.74   &   15    &  13     & 42  \\
$32^3 \times 64$ & single   &  420      &   1600     & 14       &  6.8    &   8     &  15     & 62  \\
$32^3 \times 64$ & double   &  2200     &   3900     & 6.5      &  25     &   15    &  7.0    & 86  \\
\hline
\end{tabular}
\caption{Timing benchmarks at $m=1$, $g=1/5$.  PC indicates preconditioning whereas
NPC is the inversion without preconditioning.  The time in seconds and the number
of iterations (iters.) is for convergence.  The criterion for
convergence is that the relative residual is less than $1 \times 10^{-6}$ for
single precision and less than $1 \times 10^{-12}$ for double precision.
The speed-up is the ratio of NPC time to PC time.  \label{invtab1}}
\end{table}

\begin{table}
\begin{tabular}{|c|c|c|c|c|c|c|}
\hline
Lattice             & Secs. &  Gflop/s & Iters. & Secs. & Gflop/s & Iters. \\
                    & single & single &  single & double & double & double \\
\hline
$8^3 \times 32$     & 0.93  & 15      & 360     & 3.5   & 7.5     & 700   \\
$16^3 \times 32$    & 4.8   & 28      & 370     & 20    & 13      & 730   \\
$32^3 \times 64$    & 220   & 14      & 540     & 900   & 6.7     & 1100  \\
\hline
\end{tabular}
\caption{Pseudofermion heatbath timing benchmarks at $m=1$, $g=0.1$,
with degree 20 rational approximation of \myref{hbeq}.  Preconditioning
is used up to $\beta_i=100$.  The number of iterations is the total
over the 20 different degrees.
\label{heattab1}}
\end{table}

The pseudofermion $\eta$ is updated by the heatbath method,
\beq
\eta = \( M^\dagger M \)^{1/8} R
\label{hbeq}
\eeq
where $R$ is a random complex Gaussian spinor.
The 1/8 power is obtained by rational approximation
so that rather than the problem \myref{theprob}, we
must solve
\beq
(M^\dagger M + \beta_i ) x = b
\label{yuier}
\eeq
for each degree $i=1,\ldots,d$.  In order to have
high accuracy we have taken $d=20$.
Applying the preconditioning matrix $M_0^{-1}$ then
yields
\beq
\( {\tilde M}^\dagger \tilde M + \beta_i M_0^{-1 \dagger} M_0^{-1} \) \tilde x = \tilde b
\eeq
We find that there is degradation of the speedup due to preconditioning
for larger values of $\beta_i$.  However for very large $\beta_i$ the
problem without preconditioning converges rapidly.  Thus we choose
a cutoff value, which turns out to be $\beta_i = 100$, above
which preconditioning is not used.  By this approach we are able
to keep the number of iterations for each of the problems \myref{yuier}
under 100 for single precision and under 200 for double precision.  As
Table \ref{heattab1} shows, the average number of iterations is
far less.  For example, at single precision on the $8^3 \times 32$ lattice,
a total of 370 conjugate gradient iterations are needed over 20 degrees,
for an average of 18.5 iterations for each problem \myref{yuier}.

In the molecular dynamics part of the rational hybrid Monte Carlo (RHMC) algorithm, the fermion
force terms also require the solution of equations of the form \myref{yuier}.
These have a performance similar to what is shown in Table \ref{heattab1}.
We find that approximately 50 steps with double precision are required
in order to get good acceptance rates for the Metropolis step that
is applied at the end of a molecular dynamics trajectory of length $0.5$ simulation time units.  We have
found that single precision is inadequate for the molecular dynamics
evolution.  It could however be used for a mixed precision conjugate gradient,
as has been promoted in \cite{Clark:2009wm}.

\subsection{Ward-Takahashi identity}
\label{wtis}
In the continuum the simplest Ward-Takahashi identity is that
\beq
\vev{F} = 0
\eeq
On the lattice this is modified due to the noninvariance
of the lattice action.  It is therefore of interest
to measure $\vev{F}$ in our simulations, which as
alluded to above use the RHMC algorithm.  Working at $ma=0.5$,
$g=0.1$ we find that sampling over 3283 configurations,
after 200 thermalization sweeps,
\beq
\frac{1}{V} \sum_x \vev{F(x)} = [ (2 \pm 3) + i (2 \pm 3) ] \times 10^{-3}
\eeq
The error estimates incorporate an autocorrelation time
of approximately 15 sweeps, for this observable.
The small value of $\vev{F}$ is consistent with the fact that the coupling is
weak and at one-loop order in perturbation theory,
nonrenormalization of the superpotential is intact.
Thus we expect corrections of the scalar potential to
start at $\ord{|g|^4} = \ord{10^{-4}}$, roughly consistent
with the size of $\vev{F}$.  Nonsupersymmetric corrections
to the scalar potential are required in order to have $\vev{F}$ nonzero.

\section{Conclusions}
We have studied the locality of the Dirac operator that appears
in the action \myref{lata} with fermion matrix \myref{fmat}.
It was seen in Fig.~\ref{rfig} that there is a long
tail on this operator, so that the localization is less
pronounced than that of the overlap operator.  We found
that the scalar self-energy at one-loop takes the
form \myref{onse}, so that there is no indication of
nonlocality in the external momentum.  Thus it may be
that the long tail of the operator is harmless.

We established an equivalence between the ``untilded'' formulation
of this lattice theory and the ``tilded'' theory, at the level
of correlation functions in the continuum limit, provided all divergences
are logarithmic.

We studied one-loop counterterms in the lattice theory.
We have found agreement with previous authors that in the quantum
continuum limit of this theory at one-loop, only wavefunction
renormalization occurs.  We have given numerical values and
find that in agreement with what appears in the literature,
$Z_\phi = Z_\chi$ but that $Z_F$ is different, in the continuum
limit at one-loop.

We described the renormalization of the bare lattice
supercurrent.  An approach to determining the linear
combination that is the symmetry current of the long distance effective
theory was outlined.  This is crucial for fine-tuning
the bare lattice action in the reweighting approach
that we are pursuing.  We showed that the invariance
of the free lattice action does not have a simple interpretation
in terms of a conserved supercurrent, due to the
fact that the derivative operators in this
formulation are not ultralocal.

We have developed code to run on graphics processing
units that are CUDA enabled, and have studied a strategy for
handling the inverse square roots of matrices that occur in
the theory through fast Fourier transform libraries within CUDA,
as well as our own improved version which runs on smaller lattices.
We have shown that preconditioning with the inverse of
the free theory matrix yields a speed-up that is quite significant
in the problems \myref{yuier} that are encountered in an RHMC sweep.

In future work we will fine tune the bare lattice action \myref{symgeac} through measurements
of indentities involving $\p_\mu S_\mu$, taking an approach that
involves reweighting and multicanonical Monte Carlo
techniques.  We have already measured $\vev{F}$ and find
it to be quite small at weak coupling.  Thus the bare
theory without any fine-tuning is a good starting point
for the search over parameter space.  An incremental
search strategy, starting at weak coupling, moving
gradually to stronger coupling, looks realistic.

\section*{Acknowledgements}
This research was supported in part by the Dept.~of Energy,
Office of Science, Office of High Energy Physics, Grant No.
DE-FG02-08ER41575 as well as Rensselaer faculty development funds.
We are grateful to members of the Boston University QUDA development
group (Ron Babbich, Richard Brower and Mike Clark) for providing
us a copy of their GPU code, which we borrowed from and used as a
template for our own GPU code.  We thank Georg Bergner and Alessandro
Feo for helpful discussions and communications.

\end{document}